\documentclass[aps,twocolumn]{revtex4}
\usepackage{amsmath,graphicx}

\newcommand{\be}{\begin{equation}}
\newcommand{\ee}{\end{equation}}
\newcommand{\bea}{\begin{eqnarray}}
\newcommand{\eea}{\end{eqnarray}}

\begin{document} 
 
\title{Loop models and their critical points}

\author{Paul Fendley}
\affiliation{ 
Department of Physics, University of Virginia, \\
Charlottesville, VA 22904-4714 USA \\  
{\tt fendley@virginia.edu} 
}

\begin{abstract} 

Loop models have been widely studied in physics and mathematics,
in problems ranging from polymers to topological quantum computation
to Schramm-Loewner evolution. I present new loop models which have
critical points described by conformal field theories. Examples
include both fully packed and dilute loop models with critical 
points described by the superconformal minimal models and the
$SU(2)_2$ WZW models. The dilute loop models are generalized to
include $SU(2)_k$ models as well.

\end{abstract} 

\maketitle

 
\vskip .5in 
\section{Introduction}

The statistical mechanics of loop models have been studied
extensively for decades. These models describe physical systems, and
also provide fundamental problems in mathematics.  Major progress
has been made recently in understanding their properties. This is especially
true when the loops are embedded in two dimensions, where the extensive 
variety of theoretical tools available have made many computations not
only exact, but rigorous as well.

In a loop model, the degrees of freedom are
one-dimensional.  These ``loops'' may branch and touch, but
are not allowed to have ends. The partition function is of
the form
\begin{equation}
Z= \sum_{\cal L} w({\cal L}) t^{L({\cal L})}\ .
\label{Zloop}
\end{equation}
This sum is made precise by defining the loops to live on the links of
some lattice. Then ${\cal L}$ labels a single loop configuration where
the loops have total length $L({\cal L})$, i.e.\ $L$ links of the
lattice are covered by a loop. The parameter $t$ is therefore a weight
per unit length, which in this paper will usually be tuned to a
special value which makes the behavior critical. Different loop models
are distinguished by the choice of the weight $w({\cal L})$. Here it
is required that $w({\cal L})$ depends only on {\em topological}
properties of a given configuration ${\cal L}$. Topological weights
can depend on the number of loops, how many times loops touch, and more
intricate properties like the number of ways of coloring a given
graph.


\begin{figure}[h] 
\begin{center} 
\includegraphics[width= .40\textwidth]{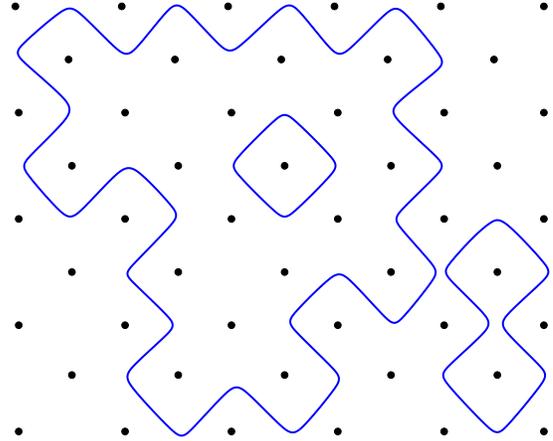} 
\caption{A typical configuration in the $O(n)$ loop model on the
  square lattice; the sites drawn are on the dual lattice.}
\label{fig:dilute-On} 
\end{center} 
\end{figure}
A simple and famous example of a loop model goes under a variety of
names: ``ring polymers'', ``closed self-avoiding random walks'', or
the ``$O(n)$ loop model'' \cite{Nienhuis}. Here the loops are defined
so that they do not branch and do not touch each other, i.e.\ if two
or more loops touch the same point, a prescription for resolving them
must be given.  A typical configuration is displayed in figure
\ref{fig:dilute-On}; the lattice is rotated by 45 degrees for later
convenience. A simple topologically-invariant quantity
is the number of loops ${\cal N}$ in a given loop
configuration ${\cal L}$.  The $O(n)$ loop model is then defined by
taking
\begin{equation}
w({\cal L})=n^{\cal N}
\label{onloops}
\end{equation}
for some parameter $n$. A single loop/walk/polymer can be obtained by 
requiring $n\to 0$. An interesting variation on this model comes from
requiring that {every} site of the lattice be visited by at least one
loop. Such loops are called {\em fully packed}, and their critical points
are very different from those of the $O(n)$
model without this requirement \cite{Nienhuis,Blote92,Kondev98}.

Many well-known models of statistical mechanics can be expressed in
terms of loops. A simple example is the Ising model in two
dimensions. A loop configuration is found from each spin configuration
simply by writing the domain walls separating regions of up and down
spins. A given loop configuration corresponds to two spin
configurations, since flipping all the spins does not change the
domain walls. These Ising loops/domain walls can touch, but do not
have ends, and cannot branch: there must be even number of links with
loops at any lattice site. The weight per unit length $t=e^{-2K}$ for
the usual Ising nearest-neighbor coupling $K=J/(k_BT)$, while the
topological weight $w({\cal L})=1$ for all allowed ${\cal L}$.

It is essential to note that statistical-mechanical models defined by
(\ref{Zloop}) do not necessarily have local Boltzmann weights. Since a
single loop can be arbitrarily large, topological properties like the
number of loops are defined non-locally. In some cases, there is a
model with local Boltzmann weights with the same partition
function. Generically, these Boltzmann weights are complex, and so the
theory is non-unitary. In some special cases like the $O(n)$
model with $n=2\cos(\pi/(k+2))$ with $k$ a positive integer, a local
model with positive Boltzmann weights does exist.

Interesting questions in both physics and mathematics arise in the
continuum limit. In this limit one can apply a variety of powerful
field-theory methods to understand the phase diagram and compute
properties at and near critical points. One major breakthrough in the
study of two-dimensional loop models came with the invention of
Coulomb-gas methods (see \cite{Nienhuis} for a review). These methods
allowed non-trivial critical points to be found, and many critical
exponents to be computed exactly, under a well-motivated set of
assumptions about the behavior of the loops in the continuum limit.
Further progress in understanding these critical points came with the
introduction of conformal field theory \cite{Belavin84}. When a
critical point in a loop model is in the same universality class as a
known conformal field theory, then not only can critical exponents be
found, but correlators computed exactly \cite{Dotsenko84}.

A remarkable recent development in the study of loop models came with
the introduction of Schramm-Loewner evolution (SLE) \cite{Cardy05}. SLE
provides a novel way of formulating conformal field theory
geometrically. Among many other things this gives a direct way of
deriving 
which conformal field theory describes certain loop models; sometimes
the correspondence can even be proved rigorously. 

Another reason for recent interest in loop models comes from the
search for quantum theories with quasiparticles obeying non-abelian
statistics. Aside from their intrinsic interest, such quasiparticles
could make up the ``qubits'' of a topological quantum computer
\cite{Kitaev97,Freedman01}. One of the few types of models 
exhibiting such quasiparticles are quantum theories whose Hilbert space has
basis elements consisting of loops in two dimensions. The ground-state
wave function is a sum over all such loops, so that ground-state
correlators in the quantum theory are identical to those in a
two-dimensional classical loop model like the ones studied here.

Despite all the recent activity and progress, not many different types
of loop models have been studied in depth. One reason is that many
loop models do not have critical points. For example, the $O(n)$ loop
model has a critical point only when $n\le 2$.  There are many
interesting and simple conformal field theories which have no known
description in terms of loops.  Moreover, for all the successes of the
SLE approach, it applies only to a very small set of conformal field
theories.

The purpose of this paper is to develop a general procedure for
finding a two-dimensional loop model with a critical point and a
corresponding conformal field theory description. This procedure
yields a number of new models whose geometric properties should be
understandable by using conformal field theory, and will hopefully
prove useful for SLE and for topological quantum computation. For
example, the loop models described here may provide a way of
geometrically interpreting the SLE processes described in
[\onlinecite{Gruzberg05}].

The strategy is to start with classical lattice models whose degrees
of freedom are not loops, but rather are {\em heights}, integer-valued
variables on the sites of the square lattice. The Boltzmann weights
are local and positive, so any field theories obtained from taking
the continuum limit of these models are unitary. The height models
utilized are {\em integrable}, which makes it possible to find a
variety of exact results. Choosing these models judiciously makes it
possible to
\begin{enumerate}
\item Locate critical
points within the lattice model, and find the
conformal field theory describing the continuum behavior of each
critical point. 

\item Re-express the partition function in terms of loop degrees of
  freedom, instead of heights, just as discussed for the Ising model above.

\end{enumerate}
Putting these two together gives a loop model whose critical point is
described by a known conformal field theory. Typically, this sort of
precise mapping will apply only when the
topological weight $w({\cal L})$ can be expressed in terms of local
quantities. However, once this map is made, the extension to
more general non-unitary cases is obvious.

There are two types of loop models to be discussed here. {\em Fully
packed} loop models have the requirement that (at least) one loop
touches every site on the lattice. Models which do not have this
requirement are referred to as {\em dilute}. The latter naming is somewhat
cavalier: the critical points of interest typically are believed to
describe a phase transition between a dilute phase (where only a set
of measure zero of the sites are touched on average by a loop), and a
dense phase (where only a set of measure zero of the sites are not
touched by a loop). Nevertheless, since at the critical point the loop
density is still zero (like the magnetization in the Ising model), I
will persist with the name. 

The dilute and fully packed loop models arise here in very different
ways. The fully packed models are found by expressing the transfer
matrix of a lattice height model in terms of the generators of an
algebra, and then finding a loop representation of the same
algebra. It is then possible to define a loop model which has the
identical partition function. The dilute models arise more indirectly:
by studying the ground states of the Hamiltonian associated with the
(same) lattice height model, one can infer what the domain walls are.
Like in the Ising model, the loops are the domain walls.  This
argument is not rigorous like that for the fully packed loops, but
substantial consistency checks are made using exact results from the
corner transfer matrix and the scattering matrix.

The outline of the rest of the paper is as follows. In section
\ref{sec:results}, I present the main results.  In
section \ref{On-cft}, I review how the ``minimal models'' of conformal
field theory are related to both dilute and fully packed loop
models. In section \ref{sec:FPL}, I introduce height models which have
critical points described by the superconformal minimal models and
$SU(2)_2$ conformal field theory. There these height models
are used to define fully packed loop models whose topological
weight is given by a chromatic polynomial. In section
\ref{sec:dilute}, dilute loop models are found which have these
critical points, and then generalized to include $SU(2)_k$ as well.
In appendix \ref{app:CFT} the relevant conformal field theories are
reviewed briefly. In appendix \ref{app:bc}, the appropriate boundary
conditions for the lattice models are discussed, while some technical
details on the BMW algebra are collected in appendix \ref{app:XE}.

\section{The loop models and the results}
\label{sec:results}

In this section I introduce the dilute and fully packed loop models
studied in this paper, and describe their critical points. The
following sections are then devoted to explaining and deriving these
results.

\subsection{Dilute loops}

\begin{figure}[h] 
\begin{center} 
\includegraphics[width= .48\textwidth]{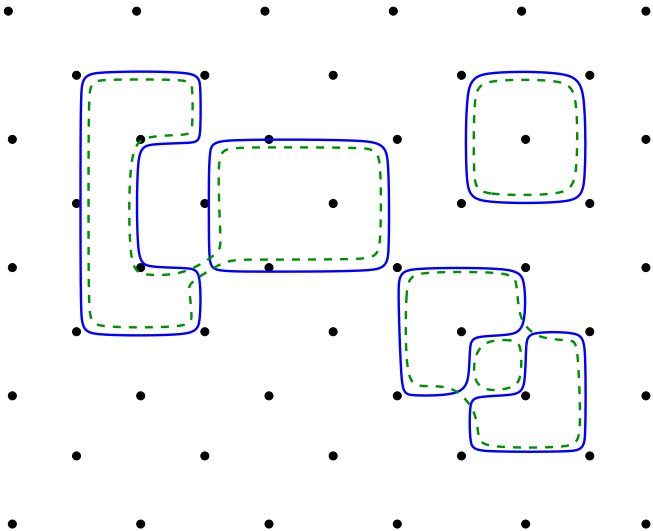} 
\caption{A typical configuration in the new dilute loop model.} 
\label{fig:dilute-k} 
\end{center} 
\end{figure}

The dilute-loop model discussed in this paper is a generalization of
the $O(n)$ loop model with two types of loops. The two types
are strongly coupled: each link containing one kind of loop must
also contain the other. A configuration is displayed in figure
\ref{fig:dilute-k}, where the two types of loops are represented by solid
and dashed lines. The loops are drawn across the diagonals of the
plaquettes of the square lattice.

In both this model and the $O(n)$ model there may be two loops coming
into the same point on the lattice, and it is required that they do
not cross.  The new model is distinguished from the $O(n)$ model by
how the loops are resolved when they potentially cross at the {\em
center} of a plaquette. In the new models, it is required that loops
of {\em each type} do not cross. Loops of different types may cross;
the four different possibilities are illustrated below in figure
\ref{fig:dilute-resolve}.

Topologically-invariant quantities of such loop configurations are the
numbers ${\cal N}$ and ${\cal M}$ of each kind of loop, and ${\cal
C}$, the number of plaquettes with a resolved potential crossing at
their center. In figure \ref{fig:FPL-BMW} these are ${\cal N}=5$,
${\cal M}=4$, and ${\cal C}=4$.  The topological weight is
\begin{equation}
w({\cal L})= n^{{\cal N}} m^{\cal M} b^{\cal C}
\label{wnm}
\end{equation}
where $n$, $m$ and $b$ are parameters.  For $m=b=1$, this weight reduces to
that of the $O(n)$ model. It is convenient to
parametrize $n$ and $m$ by
\begin{eqnarray}
\nonumber
n&=& 2 \cos\left(\frac{\pi}{p-k+1}\right)\ ,\\
m&=& 2 \cos\left(\frac{\pi}{k+2}\right)\ .
\label{nm}
\end{eqnarray}

I will argue that with appropriate boundary conditions, these dilute
loop models have a critical point described by a unitary conformal
field theory when $p$ and $k$ are integers with $p>k+1$ and
$k>1$. These conformal field theories are called coset models, and are
described briefly in appendix \ref{app:CFT}. The $k=1$ case reduces to
the $O(n)$ model, which is well known to be described by the conformal
minimal models \cite{Dotsenko84}. For $k=2$, the conformal field
theories are the superconformal minimal models, whereas for
$p\to\infty$ ($n\to 2$), they are the $SU(2)_k$ Wess-Zumino-Witten
models. It is natural to conjecture that this loop model
has a critical point when $n\le 2$ and $m\le 2$, although for $p$
and $k$ outside these special values, the corresponding 
conformal field theories will be non-unitary.

\subsection{Fully packed loops}

\begin{figure}[h] 
\begin{center} 
\includegraphics[width= .43\textwidth]{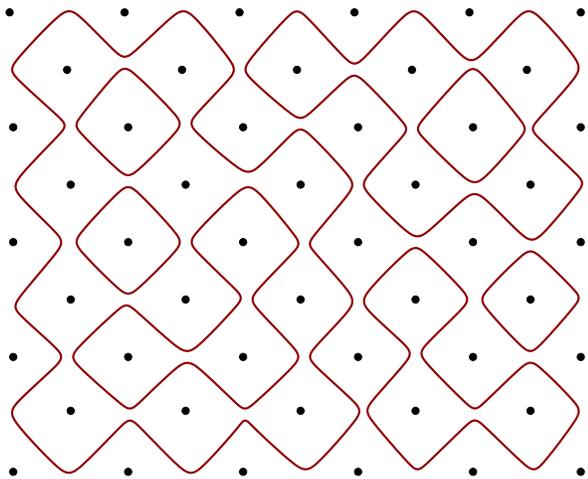}
\caption{A typical configuration in a fully packed $O(n)$ loop model} 
\label{fig:FPL-On} 
\end{center} 
\end{figure}
A fully packed version of the $O(n)$ loop model is illustrated in
figure \ref{fig:FPL-On}. In this version the loops cover every link of
the square lattice. The only degree of freedom is how the loops are
resolved at each site of the lattice: there are two possible
resolutions which do not allow crossings. The topological weight is
then
\begin{equation}
w_F({\cal L})= (\sqrt{Q})^{\cal N}
\label{wF}
\end{equation}
when there are ${\cal N}$ of these loops. The weight per loop is
labeled $\sqrt{Q}$ here because for weight per unit length $t=1$ and
appropriate boundary conditions, this model is equivalent to the
${Q}$-state Potts model at its self-dual point. This is shown using
the high-temperature expansion of the Potts model \cite{FK,Baxbook},
which will be reviewed in section \ref{sec:TL-RSOS}. The Potts
self-dual point is critical when ${Q}\le 4$, so that the critical
point occurs for a weight per loop $\le 2$, just like the dilute
$O(n)$ models. Also like the dilute $O(n)$ models, the continuum limit
of the critical Potts models are described by the minimal models of
conformal field theory when ${Q}$ takes on special values. One
obtains the $p$th minimal model when $\sqrt{Q}= 2\cos(\pi/(p+1))$
with $p$ integer; note the shift of $p$ relative to the dilute value
found in (\ref{nm}) with $k=1$. This critical point is often called
the {\em dense} critical point of the $O(n)$ loop model
\cite{Nienhuis,Blote92}.

\begin{figure}[h] 
\begin{center} 
\includegraphics[width= .43\textwidth]{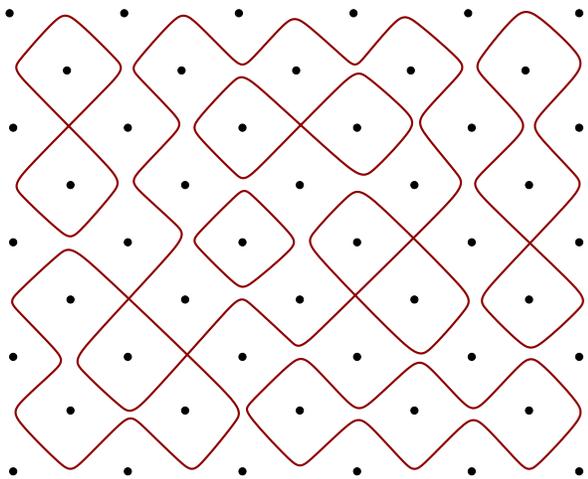}
\caption{A typical configuration in the fully packed $k=2$ model} 
\label{fig:FPL-BMW} 
\end{center} 
\end{figure}
In terms of the more general coset conformal field theories with
central charge (\ref{ccoset}), the minimal models and hence the dense
critical points have $k=1$. In section \ref{sec:FPL} I derive fully
packed loop models which have critical points corresponding to the
$k=2$ case, the superconformal minimal models. A typical configuration
in this fully packed model is illustrated in figure
\ref{fig:FPL-BMW}. There are three allowed configurations for each
vertex, the two ways of resolving the lines so that they do not cross,
and a third, which I call the ``intersection''. Since these
intersections are not resolved, the configurations do not really form
loops, but for lack of a better name, I still call the degrees of
freedom loops.

\begin{figure}[h] 
\begin{center} 
\includegraphics[width= .3\textwidth]{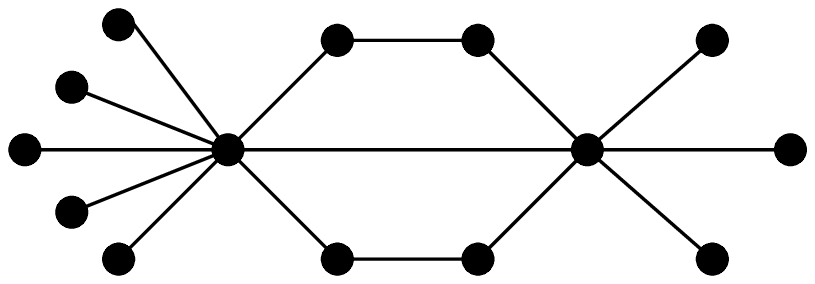} 
\caption{The dual graph for the loop configuration in figure
\ref{fig:FPL-BMW}}
\label{fig:dualgraph} 
\end{center} 
\end{figure}

The topological weight of the model is not defined by resolving the
intersections to make the configurations into non-crossing
loops. Instead, it is given in terms of the {\em chromatic
polynomial}.  To define the chromatic polynomial, it is first useful
to note that the loops divide the plane into different regions. These
regions can be thought of as countries on a map, and a $Q$-coloring
gives each country one of $Q$ colors, such that adjacent countries
must have different colors. If two regions touch only at a point (or
any countable set of points), they are not adjacent. This is
convenient to formulate in terms of the dual graph. Each region
corresponds to a vertex on the dual graph, and two vertices are joined
by an edge if the corresponding regions are adjacent. The dual graph
for the loop configuration in figure \ref{fig:FPL-BMW} is illustrated
in \ref{fig:dualgraph}.  Note that the dual graph only depends on the
topology of the original loop configuration.  The number of
$Q$-colorings $\chi_Q({\cal L})$ of the dual graph is the number of
ways of assigning $Q$ colors to each vertex with the constraint that
any vertices joined by an edge must be a different color.


There exists no
closed-form formula for the number of colorings of a general
graph. However, $\chi_Q$ can be determined recursively.  Consider
two vertices connected by an edge $l$ (i.e.\ two regions sharing
a boundary in the original picture). Then define ${\cal D}_l{\cal L}$
to be the graph with the $l$ edge deleted, and ${\cal C}_l{\cal L}$ to be
the graph with the two vertices connected by $l$ joined into one. Then
it is simple to prove that
\begin{equation}
\chi_Q^{}({\cal L}) = \chi_Q^{}({\cal D}_l{\cal L}) - \chi_Q^{}({\cal C}_l{\cal L}).
\label{recursion}
\end{equation}
In the dual graph of ${\cal L}$ the two vertices connected by $l$ must
be colored differently, while in ${\cal D}_l{\cal L}$, they may be the same,
Thus $\chi_Q({\cal D}_l{\cal L})\ge \chi_Q({\cal L})$, and the
number of colorings overcounted is $\chi_Q({\cal C}_l{\cal L})$.
The number of colorings can then be determined by using this recursion
relation to remove edges and vertices from ${\cal L}$ until it is a
sum of terms which have no edges at all. A term with ${\cal
V}$ vertices and no edges has $\chi_Q=Q^{\cal V}$. An illuminating
example is the case where there are no intersections, so that the
configurations really are loops. The dual graph is then tree-like (it
has no faces). Vertices at the ends of the tree can be removed
one by one using (\ref{recursion}), with each vertex being removed
giving a factor $Q-1$. This yields $\chi_Q=Q(Q-1)^{{\cal V}-1}$ for
a tree with ${\cal V}$ vertices.  Using the
recursion relation (\ref{recursion}) gives for the loop configuration in
figure \ref{fig:FPL-BMW} and the dual graph in figure
\ref{fig:dualgraph}
$$\chi_Q= Q(Q-1)^{9}(Q^2-3Q+3)^2.$$

Repeatedly applying the recursion relation gives $\chi_Q({\cal
L})$ as a polynomial in $Q$ of order ${\cal V}$. This is known as the
chromatic polynomial, and depends only on the dual graph of ${\cal
L}$. This polynomial provides a well-defined extension of $\chi_Q$
away from $Q$ integer.  The fully packed loop model is therefore
uniquely defined for any $Q$ with topological weight
\begin{equation}
w({\cal L})=\chi_Q({\cal L}) (\sqrt{Q}+1)^{{-\cal N}_X}.
\label{wchrome}
\end{equation}
${\cal N}_X$ is the number of  intersections; the factor involving it arises from the detailed
analysis below. The full
partition function is given by using this topological weight with
(\ref{Zloops}), where the weight per unit length of loop is $t=1$.
The sum is over all loops on the square lattice such that each
plaquette has one of the three configurations drawn in figure
\ref{fig:IXE}.

In section \ref{FPLk2} I prove that the loop model with
topological weight (\ref{wchrome}) has a critical point described by
the $p$th superconformal minimal model (i.e.\ those in (\ref{coset})
with $k=2$), when
\begin{equation}
Q= 4\cos^2\left(\frac{\pi}{p+1}\right)\ .
\end{equation}
These fully packed models presumably remain critical for all real
$p$, but are described by a non-unitary conformal field theory for $p$
non-integer.

This loop model is very similar to that of \cite{Fendley05}, which
arises from the low-temperature expansion of the $Q$-state Potts
model.  The topological weight is given by the chromatic
polynomial. However, the loops there are dilute, not fully packed, and
intersections with three lines coming out are allowed there. As a
result, the critical point there is described by the conformal minimal
models instead of the superconformal minimal models.

This is not the only fully packed loop model which has critical points
corresponding to the superconformal minimal models. One based on
non-intersecting loops on the ``copper-oxide'' lattice will also be
briefly discussed below.

\section{The $O(n)$ loop models}
\label{On-cft}

The strategy of this paper is to use known integrable lattice models
to find loop models with critical points described by conformal field
theories. In this section this strategy is used to connect the $O(n)$
loop models with the minimal models of conformal field theory. None of
the results in this section are new, but possibly some of the
arguments are.  After introducing the lattice models, the two
different possible loop expansions are derived. Even though both loop
expansions come from the same lattice model, they arise in very
different ways.  The fully packed loops arise by rewriting the model
in terms of an algebraic formulation of the transfer matrix, which in
most contexts is a high-temperature expansion. Dilute loops arise from
a domain-wall expansion, which is a low-temperature expansion.

\subsection{RSOS lattice models}
\label{sec:TL-RSOS}

The lattice models here are the ``restricted-solid-on-solid'' (RSOS)
models introduced by Andrews, Baxter and Forrester \cite{ABF}. The
degrees of freedom are integer-valued heights $h_i$ on each site $i$
of the square lattice. There are two types of restriction. The first
is that $h_i$ only takes a finite number of possibilities, which are
numbered from $1\dots p$ for some integer $p$. The second is that
heights on adjacent sites must differ only by 1, i.e. $|h_i-h_j|=1$
for $i$ and $j$ nearest neighbors. The latter restriction will be
modified for the models in later sections. In order to utilize crucial exact
results, I study only height and loop models on the square lattice. However,
frequently (but not always), one obtains similar results on different
two-dimensional lattices.

The $p=3$ case is
the Ising model: on one sublattice the height must always be $2$, so
that on the other sublattice the heights $1$ and $3$ play the roles of
the $+$ and $-$ Ising spins. The $p=4$ RSOS model is known as the hard
square model \cite{Baxbook}; a typical configuration is illustrated in
figure \ref{fig:hardsq} below.

The RSOS models are interesting and useful because for special choices of the
Boltzmann weights, they are integrable. Integrability allows a number
of important physical quantities to be computed exactly, under a set
of standard analyticity assumptions. Critical points and the
associated exponents can be found using a powerful technique
called the corner transfer matrix \cite{Baxbook}. Conformal field
theory also provides a list of critical exponents associated with
critical points. By comparing the two lists, 
one can usually identify the conformal field theory describing the
continuum limit of any critical point in an RSOS model. 
For these RSOS models, there are
two known critical points for each value of $p>3$.  This paper is 
concerned with just one of them, the one separating regimes III and IV
in the nomenclature of [\onlinecite{ABF}]. This critical point is 
described by the $p$th minimal model, with central charge given by
(\ref{cmin}) \cite{Huse}.

\begin{figure}[h!]
\begin{center}
\includegraphics[width= .48\textwidth]{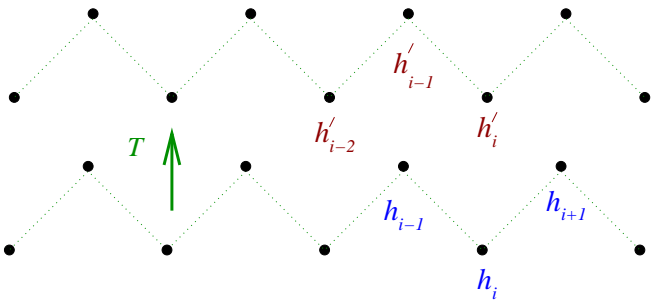} 
\caption{How the transfer matrix acts}
\label{fig:diag}
\end{center}
\end{figure}

Finding the loop models associated with this RSOS model and hence the
conformal minimal models requires analyzing the Boltzmann weights in
detail. The weights contains nearest-neighbor and
next-nearest-neighbor interactions, so the total weight can be written
as a product of weights assigned to the four heights around each
square plaquette. Such a model is called ``interaction round a face''
\cite{Baxbook}.  The transfer matrix $T_{h,h'}$ is defined to act
across the diagonals of the lattice, as illustrated in figure
\ref{fig:diag}.  
It therefore acts on a zig-zag row of heights $h_i$,
$i=0\dots 2N$, taking it to the next row, labeled by $h_i'$. It can be
broken into pieces ${\cal T}_i$, which depend only on the heights
around a square, and can change $h_i$ to $h'_i$. Then $T_{h,h'}$ can
be written in the form
\begin{equation}
T=  
{\cal T}_{1}{\cal T}_3\ \dots\ {\cal T}_{2N-1}
I_0 {\cal T}_2{\cal T}_4\ \dots\ {\cal T}_{2N-2} I_{2N}
\label{Tprod}
\end{equation} 
where  the identity matrix $I_i =\delta_{h_i
  h'_i}$. For the RSOS critical point of interest here, we have
\begin{equation}
{\cal T}_i = I_i + x e_i
\label{Te}
\end{equation}
where $x$ is a parameter, and the elements of the matrix $e_i$ for $i$ even
are \cite{ABF,Pasquier87a}
\begin{equation}
e_i = \delta_{h_{i-1} h_{i+1}} \frac{ \sqrt{[h_i]_q
    [h'_i]_q}}{[h_{i+1}]_q} \prod_{j\hbox{ even, }j\ne i} I_j
\label{eRSOS}
\end{equation}
where 
$$[h]_q\equiv (q^h - q^{-h})/(q-q^{-1}),\qquad  q\equiv e^{\pi i/(p+1)}.$$
Acting with $e_i$ therefore 
allows the height on the $i$th site to change when
$h_{i-1}=h_{i+1}$.
For $i$ odd, the matrix $e_i$ is essentially the
same:
\begin{equation}
e_i = \delta_{h_{i-1}' h_{i+1}'} \frac{ \sqrt{[h_i]_q
    [h'_i]_q}}{[h'_{i+1}]_q} \prod_{j\hbox{ odd, }j\ne i} I_j
\label{eRSOSii}
\end{equation}

The RSOS transfer matrix written in this form 
exhibits an interesting and important algebraic structure: the
$e_i$ generate the {\em Temperley-Lieb
  algebra} \cite{TL}. It is straightforward to check that they satisfy
\begin{eqnarray}
e_i^2 &=& (q+q^{-1}) e_i,\nonumber\\ e_i\, e_{i\pm 1}\, e_i &=&
e_i,\nonumber\\  e_i\,e_j&=&e_j\,e_i \quad (|j-i|\ge 2).
\label{TLalg}
\end{eqnarray} %
This algebra first arose in studies of the $Q$-state Potts model,
whose transfer matrix also can be written in the
form (\ref{Tprod},\ref{Te}), with the $e_i$ satisfying the same algebra
(\ref{TLalg}) with $q+q^{-1}=\sqrt{Q}$. Since then, it
has been shown how to write many other lattice models in this same
form. 

Writing a transfer matrix in terms of the Temperley-Lieb generators is
exceptionally useful because many properties of the model follow
solely from the algebra, and are independent of the representation of
the $e_i$. With appropriately-chosen boundary conditions, the
partition function is independent of representation. This is shown for
a two-dimensional surface with the topology of a sphere in the
appendix \ref{app:bc}, while detailed discussions for the cylinder and
torus can be found in [\onlinecite{Pasquier87b,DSZ87}]. In many considerations,
understanding the boundary conditions precisely is very important; for
example, the central charge of the continuum conformal field theory
depends on them. For the results here, the details of the
boundary conditions are not particularly important -- what is of main
importance is that appropriate ones exist.

\subsection{Fully packed loops}
\label{TL-FPL}

This construction of the fully packed loop model associated with each
critical RSOS model is virtually identical to that of the
fully packed loop model for the Potts model \cite{FK,Baxbook}. 
This is easiest to do by using a
graphical representation of the Temperley-Lieb
generators \cite{Pasquier87b}.

\begin{figure}[h] 
\begin{center} 
\includegraphics[width= .25\textwidth]{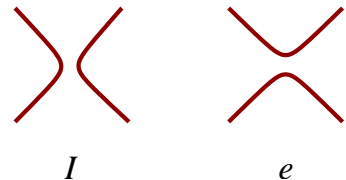} 
\caption{The graphical representation of the Temperley-Lieb generators} 
\label{fig:TL} 
\end{center} 
\end{figure}

Fully packed loop configurations are not in one-to-one correspondence
with those of the height or Potts model. Rather, for each height
configuration, the product in (\ref{Tprod}) is expanded out by using
(\ref{Te}). One obtains either the identity $I$ or the
Temperley-Lieb generator $e$ for each plaquette, so that for ${\cal P}$
plaquettes there are $2^{\cal P}$ terms. Each term can be graphically
represented by drawing one of the two pictures in figure \ref{fig:TL}
for each plaquette, depending on whether $I$ or $e$ appeared for this
plaquette.  This gives a fully packed configuration of self-avoiding
and mutually-avoiding loops on the dual square lattice. Every link of
the dual lattice is covered by a loop, and at each vertex there are
two possible ways of resolving the lines so that the loops do
not cross. These properties can be preserved at the boundaries as
well, as described in appendix \ref{app:bc}.

\begin{figure}[h] 
\begin{center} 
\includegraphics[width= .48\textwidth]{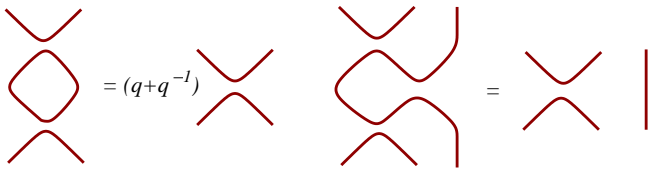} 
\caption{The graphical representation of the Temperley-Lieb algebra} 
\label{fig:TLalg} 
\end{center} 
\end{figure}

To obtain the fully packed loop model, one must sum over the heights
in the RSOS representation (or the spins in the Potts representation),
leaving only the loops themselves as the degrees of freedom. The sum is
not difficult to do directly in the RSOS representation
\cite{Pasquier87b} (or in the Potts representation \cite{FK,Baxbook}),
by using the explicit representation for $e$, as given in
(\ref{eRSOS}). It is more instructive, however, to not utilize the
details of any explicit representation, but rather to exploit the
properties of the Temperley-Lieb algebra. In the transfer matrix,
summing over heights amounts to multiplying the matrices defined by
$I_i$ and $e_i$. Therefore, the algebra itself can be represented
graphically as in figure \ref{fig:TLalg}. From this it is apparent
that small closed loops receive a weight $q+q^{-1}$.  By using both
relations, it is easy to check that {\em all} loops receive this same
weight $q+q^{-1}$, no matter what their size (for details, see
appendix \ref{app:bc}). Setting $x=1$ to make the model isotropic
(invariant under 90-degree rotations) yields the fully packed $O(n)$
loop model discussed in the introduction. The weight per loop is
$n=q+q^{-1}=2\cos[\pi/(p+1)]$ for the RSOS model, or $n=\sqrt{Q}$ for
the $Q$-state Potts model. The weight per unit length of loop $t$ in
this case is unimportant, because every link is covered by a loop.

The $O(n)$ loop model can of course be defined for any $n$, but is
only critical for $|n|\le 2$. This is shown by using
the representation of yet another lattice model, the six-vertex model,
in terms of the Temperley-Lieb algebra \cite{Baxbook}. Another
interesting thing to note is that even when $|n|\le 2$, the conformal
field theory is unitary only for integer $p\ge 3$. 
The representation (\ref{eRSOS}) exists for any value
of $q=e^{i\pi/(p+1)}$, but the only ones leading to real and positive
Boltzmann weights for $q+q^{-1}\le 2$ occur at $p$ integer. 

\subsection{Domain walls in the RSOS model}
\label{TL-dilute}

The dilute $O(n)$ loop model also has a critical point related to the
conformal minimal models \cite{Dotsenko84}. A way of showing this
heuristically is to compare the Coulomb-gas description of the minimal
models \cite{Dotsenko84,Dotsenko84b} to that arising from the lattice
model \cite{Nienhuis}. A way of showing this directly is to study a
different series of integrable RSOS models known as the ``dilute
Temperley-Lieb'' models \cite{Nienhuis90,Warnaar93a,Warnaar93b}. In this
section I give another method of relating the two, by using the RSOS
models defined in section \ref{sec:TL-RSOS}. 
This is the method which readily generalizes to $k>1$.

It is useful to start by discussing how the dilute
Temperley-Lieb models are related to the $O(n)$ loop model
\cite{Nienhuis90,Warnaar93a}.  The dilute models are defined in terms of
integer heights on the square lattice like the RSOS models discussed
above in section \ref{sec:TL-RSOS}, but the restriction on neighboring
heights in the dilute models is relaxed to $h_i-h_j=0,\pm 1$.  The
loops are simply the domain walls between different heights. The
``dilute'' in the name refers to the fact that because $h_i-h_j$ can
vanish for nearest neighbors, there need not be domain walls on every
link. Just drawing domain walls does not immediately make it into the
$O(n)$ loop model: there is no reason {\it a priori} why a region of 
heights $1$ and $2$ has the same weight
as a region of heights $2$ and $3$. Nevertheless, when
the Boltzmann weights are tuned appropriately, one can perform the sum
over heights to leave a model consisting of self-avoiding and
mutually-avoiding loops with weight $n$ per loop \cite{Nienhuis90}. 
By then doing a corner transfer matrix computation, one finds indeed that the
continuum limit is described by a conformal minimal model, with the
weight $n$ per loop related to $p$ by $n=2\cos(\pi/p)$ (note the shift
in $p$ as compared to the fully packed case) \cite{Warnaar93b}.

The loops discussed in the remainder of this section can be thought of
as domain walls for the RSOS model defined in section
\ref{sec:TL-RSOS}. However, they are not simply domain walls between
differing heights, as they are in the dilute model. Instead, they are
best thought of as domain walls between ``ground states'' in a
1$+$1-dimensional picture.  Namely, one chooses a Euclidean time
direction, and takes the continuum limit in the time direction to
obtain a one-dimensional quantum Hamiltonian, and find its ground
states. Excitations in the one-dimensional picture are then kinks, or
defects, which separate regions comprised of the different ground
states. The world lines of the kinks in 1$+$1 dimensions can be
thought of as domain walls in the
two-dimensional classical model. Obviously, this picture is heuristic,
but below I will explain how to use it to develop exact results for
the RSOS model into a precise conjecture for a dilute loop model.

A quantum Hamiltonian for the RSOS models is obtained in the limit
$x\to 0$. One finds that the critical properties of the model are
independent of $x$, as along as its sign is not changed. Thus taking
$x\to 0$ in (\ref{Tprod}) and (\ref{Te}) gives a quantum Hamiltonian
acting in the same direction as the transfer matrix, i.e.\ across the
diagonals of the lattice. It is (for $x>0$)
\begin{equation}
H= -\sum_{i=1}^{2N-1} e_i\ .
\label{He}
\end{equation}
This Hamiltonian acts on the  ``height chain'' $(h_0,h_2,\dots,h_{2N})$.

Using (\ref{eRSOS}), it is easy to find candidates for the ground states
of (\ref{He}). The matrix elements of $e_i$ are positive when
$h_{i-1}=h_{i+1}$, and vanish when $h_{i-1}\ne h_{i+1}$. Thus it is
natural to expect that a ground state will be dominated by states that
obey $h_{i+1}=h_{i-1}$ for all $i$. The RSOS models defined in section
\ref{sec:TL-RSOS} have the restriction that $h_{i}-h_{i-1}=\pm 1$ for all
$i$. Thus potential ground states here are of the form
$(\dots r,r+1,r,r+1,\dots)$.

Finding which one or ones of these potential ground states dominates
the actual ground state or states is in general an imposing
problem. For the integrable Hamiltonian (\ref{He}), however, it is
possible to answer this question by using the corner transfer matrix
technique \cite{Baxbook,ABF}. The result is that there is a ground
state dominated by {\em each} of the potential ground states.
Thus this critical point is a multicritical point, where all the
potential orderings coexist \cite{Huse}.

I illustrate this here in the simplest cases, and then in the
next subsection \ref{TL-CTM} explain
how the general result follows from the corner transfer matrix. The
simplest case is the Ising model, which has $p=3$. The restriction
that $|h_i-h_{i+1}|=1$ means that every other site on the chain (say
the even sites) must be occupied by height $2$. The odd sites are
occupied by heights $1$ and $3$, which play the role of the spins
$\pm$ in the usual formulation of the Ising chain. Using the explicit
matrix elements given in (\ref{eRSOS}) means that $-e_i$ for
even $i=2j$ yields a potential energy which is $-\sqrt{2}$ if
$h_{2j-1}=h_{2j+1}$, and $0$ for $h_{2j-1}\ne h_{2j+1}$. For odd $i=2j-1$,
$$e_{2j-1}=\frac{1}{\sqrt{2}}
\begin{pmatrix}
1&1\\
1&1
\end{pmatrix}\ ,
$$
where the rows and columns are indexed by $h'_{2j-1}=1,3$ and $h_{2j-1}=1,3$
respectively. Rewriting the Hamiltonian in terms of the Pauli matrices
$\sigma^a(j)$ gives
$$H_{p=3} = -\frac{1}{\sqrt{2}}\sum_{j=1}^{N/2} 
\left[ \sigma^z(j)\sigma^z(j+1) +t \sigma^x(j)\right]
$$ where $t=1$. This is exactly the ferromagnetic Ising spin
chain; $t=1$ is the critical point. For $t<1$, the chain
is ordered: there are two ground states, one dominated by the
configuration with by all spins up, the other dominated by the
configuration with all spins down. In the height language, these
correspond respectively to the configurations ($121212\dots$) and
($323232\dots$). For $t>1$, the model is disordered. Thus
$t=1$ is the well-known Ising critical point between the ordered
and disordered phases. Excited states in the ordered phase contain
defects, i.e.\ states which contain regions of both ground states
locally. The equivalence of the Ising model to the dilute loop model
with $n=2(\cos(\pi/3))=1$ was already discussed in the
introduction. In the height language, the loops are domain walls
separating regions of height $1$ and height $3$.

The two ground states in the $p=3$ case are symmetry-equivalent: the
critical point is just the usual one where the discrete symmetry
becomes no longer spontaneously broken. For $p>3$, the different
ground states
are no longer all related by any obvious symmetry, but the fine tuning
necessary to get to the integrable multi-critical point with
Hamiltonian (\ref{He}) makes them all
degenerate.

\begin{figure}
\includegraphics[width= .42\textwidth]{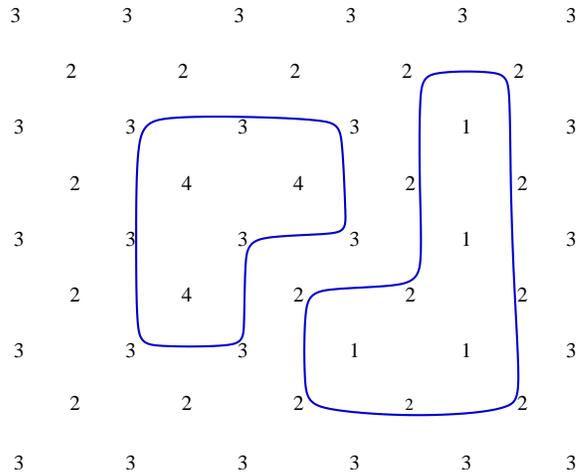} 
\caption{A typical configuration in the RSOS height model with $p=4$. The
  lines are the domain walls corresponding to dilute loops, drawn so
  that they do not cover the heights that they go through.}
\label{fig:hardsq}
\end{figure}

To see this in more detail, consider the $p=4$ case, the
``hard-square'' model \cite{Baxbook}. The hard-square tiles have diagonal
length twice the lattice spacing, and are placed with their centers on
the sites of a square lattice,
with the rule that no two tiles can overlap. This then forbids
tiles from being adjacent to each other. In the height language,
this means that heights $1$ and $4$ correspond to the tiles, and $2$
and $3$ correspond to empty sites (on one sublattice the height is
always even, on the other it is always odd, and the Boltzmann weights are
invariant under $h\to p+1-h$.) The restriction that
$|h_i-h_j|=1$ for $i$ next to $j$ enforces the hard-square
restriction. A typical configuration is displayed in figure \ref{fig:hardsq}. 
The Hamiltonian is easiest to write by treating the
square tile as a hard-core boson created by $d^\dagger_i$. It is then
\cite{Fendley04}
\begin{multline}
\mathcal{H} = \sum_i \left[ -W \left( d_i + d^{\dagger}_i \right)
(1-n_{i-1})(1-n_{i+1}) \right.\\
\left.
+ U n_i + V n_i n_{i+2} \right] 
\label{hamsquare}
\end{multline}
where $n_i=d^\dagger_id_i$ is the number operator at site $i$.
The critical (\ref{He}) corresponds to the values
\begin{equation}
W_c=\phi^{-1/2}, \qquad V_c=-\phi^2, \qquad U_c=2+1/\phi^2
\label{wVU}
\end{equation}
 where
$\phi=2\cos(\pi/5)=(1+\sqrt{5})/2$ is the golden mean.


To understand the multiple ground states, it is useful to first take
the limit where $|U/W|$ and $|V/W|$ are large, so that the
off-diagonal term proportional to $W$ can be neglected. For $U>-V$
then the ground state is the state with no bosons, while for $U<-V$
there are two ground states, each of which has a particle on every other
site. In the latter regime, a ${\bf Z}_2$ symmetry (translation by one
site) is spontaneously broken. At $U=-V$, there is a first-order phase
transition between the ${\bf Z}_2$-broken phase to an unbroken
phase. In height language the two ground states in the ${\bf
Z}_2$-broken phase are dominated by the states $(\dots 12121212\dots)$
and $(\dots 34343434\dots)$ respectively, while the single ground
state in the unbroken phase (the empty state in the bosonic language)
is dominated by $(\dots 32323232\dots)$.  This first-order transition
persists as $W$ is included; doing perturbation theory 
gives the location of this transition at next order to be
\cite{Fendley04}
\begin{equation}
\frac{U}{V} = -1+ \frac{W^2}{V^2},
\label{firstorder}
\end{equation}
Although the perturbative computation is only reliable to order
$(W/V)^2$, it turns out that (\ref{firstorder}) is the exact
first-order transition line for $U>U_c$. Along this line the two
phases coexist -- the hard-square model has three symmetry-equivalent
ground states.  When the couplings are tuned to the values in
(\ref{wVU}), this first-order line terminates in the tricritical point
\cite{Baxbook,Huse83}.


This information about the ground states is sufficient to guess what
the lowest-energy excitations are; in the next subsection \ref{TL-CTM}
I will justify these guesses using the corner transfer matrix. Excited
states along this first-order transition line are defects, where two
different ground states meet. For example, the state $(\dots
32323232121212323232 \dots)$ has two defects, and energy order $U$
higher than the ground state. Thus each of these defects has energy of
order $U/2$. Such a defect occurs at site $i$ any time
$|h_{i-1}-h_{i+1}|=2$.  There are two kinds of fundamental defects,
with the same energy.  Denoting the three ground states as $G_{12}$,
$G_{23}$ and $G_{34}$, where $G_{ab}=(\dots abababab\dots)$, one kind
of fundamental defect separates $G_{12}$ and $G_{23}$, while the other
separates $G_{23}$ and $G_{34}$. The former defect is located at the
site where there is a height $2$ between the heights $1$ and $3$,
while the latter is located at the site where there is a height $3$
between heights $2$ and $4$. A defect $(\dots 1212343434 \dots)$
separating $G_{12}$ from
$G_{34}$ has twice the energy as the
fundamental defects, and is comprised of two fundamental defects.

These statements can be translated from the one-dimensional quantum
Hamiltonian back to the two-dimensional hard-square model
\cite{Huse83}. The three ground states of the Hamiltonian result in
three different ways the {\em free energy} density can be
minimized. One can easily check that at the critical point, the weight
of the configuration with all heights $1$ and $2$ (which in a slight
abuse of notation can be labeled $G_{12}$) is larger than that of the
configuration containing all heights $2$ and $3$ (the 2d analog of
$G_{23}$). On the first-order line, there are three
different ways the free-energy density can be minimized. The reason is
that there are more configurations ``close'' to $G_{23}$ than there
are to $G_{12}$ or $G_{34}$, so the increased entropy will compensate
for the increased energy. This is the cause as well of the shift of
the first-order line in (\ref{firstorder}). With $G_{23}$, one can
change any height $2$ to height $4$, and any height $3$ to height $1$,
without violating the hard-square rule. With $G_{12}$, one can change
any height $1$ to height $3$, but it is not possible to change any
heights $2$ to height $4$ without violating the restriction, unless
one also changes the four heights $1$ around it. In the language of
hard squares, there are more ways of adding a square to the empty
configuration than there are of removing one from the full
configuration, since in the latter the squares occupy only every other
site. Thus the three ground states turn into the three minima of the
free energy.

The partition function can therefore be rewritten as a sum over
domain-wall configurations separating the free-energy minima.
The excited states of the Hamiltonian become the domain
walls in the 2d classical model, separating regions of ${\bf Z}_2$
broken and unbroken symmetry. These domain walls are the world-lines
of the one-dimensional defects, and their contribution to the free energy
depends only on their length times some positive constant $f_D$. By
the symmetry $h\to p+1-h$, $f_D$ must be the same for both kinds of
domain wall. A
non-vanishing $f_D$ is the reason the transition is first-order: the
tricritical point occurs when the couplings are tuned to make $f_D\to
0$. In this limit there is no free-energy penalty for creating a domain wall,
and they proliferate. 

It is finally possible to give the argument as to why the hard-square
model is equivalent to a loop model, in the sense that they are
described by the same field theories in the continuum limit. I have
argued that the hard-square model along its first-order line can be
thought effectively as having three free-energy minima.  Domain walls
separating regions of these minima have a free-energy weight per unit
length. The domain walls form loops which do not touch: as noted
above, domain walls separating $G_{12}$ and $G_{23}$ are located along
sites with height $2$, while domain walls separating regions $G_{23}$
and $G_{34}$ are located along sites with height $3$.  These domain
walls are illustrated in figure \ref{fig:hardsq}.  I emphasize that
this picture involving free energies and domain walls is effective:
these should be thought of as renormalized quantities.

Nevertheless, these results imply that along the
first-order line, the partition function of the hard-square model can
be written in the form
\begin{equation}
Z_{p=4}\approx \sum_{\cal L} w({\cal L}) e^{-f_D L}.
\label{Zloops}
\end{equation}
${\cal L}$ is a configuration of loops of total length $L$ on the
links of the dual square lattice, with the condition that they cannot
touch. In this expansion, no distinction is made between the two
different types of domain walls: the fact there are two types of loops
can be taken into account in the weight $w({\cal L})$. To do
this, draw a
loop configuration. The restrictions on heights mean that a loop
separates a region of $G_{23}$ from a region of either $G_{12}$ or
$G_{34}$. Since the energies of $G_{34}$ and $G_{12}$ are related by
the ${\bf Z}_2$ symmetry, we can then sum over these two
possibilities. Half the regions inside the loops therefore must be
$G_{23}$, while the other half can be either $G_{12}$ or
$G_{34}$. This results in a weight $w({\cal L})=2^{{\cal N}/2}$, where
${\cal N}$ is the total number of loops, so that the weight per loop
is $\sqrt{2}$. This is the dilute $O(n)$ loop model with
$n=\sqrt{2}$. The (tri)critical point occurs at $f_D=0$ where
there is no penalty for longer loops and they can proliferate. 

The loop model therefore arises here indirectly, as
opposed to the dilute Temperley-Lieb models, where the domain walls
and hence the loops are manifest in the original lattice height
configurations \cite{Nienhuis90,Warnaar93a}. All arguments give the
same result: the conformal field theory with $p=4$ describes the
critical point of the $O(\sqrt{2})$ dilute loop model on the sphere.

\subsection{Dilute loops and the corner transfer matrix}
\label{TL-CTM}

In the preceding subsection \ref{TL-dilute} I detailed an argument
relating the RSOS model with $p=4$ to the $O(\sqrt{2})$ loop
model. The argument can be strengthened and generalized to arbitrary
$p$ by using the corner transfer matrix technique \cite{Baxbook,ABF}.
The corner transfer matrix allows one to find the ground states in
general, and also gives essential information about the defect
energies.

\begin{figure}[h] 
\begin{center} 
\includegraphics[width= .48\textwidth]{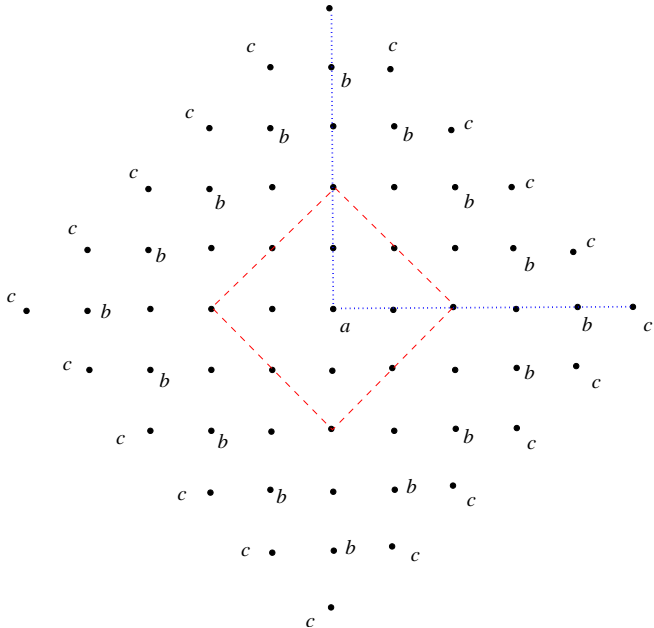} 
\caption{The corner transfer matrix computes the partition function in
  one quadrant of the square. In the frozen limit, the spins are all
  the same along a given diagonal of each quadrant.}
\label{fig:CTM} 
\end{center} 
\end{figure}
Consider a two-dimensional lattice model on a square lattice, as
illustrated in figure \ref{fig:CTM}.
The corner transfer matrix acts on $N$ sites
along a line, taking them to $N$ sites along a line {\em
  perpendicular} to the original line. The partition function for the
full model is then built up from four corner transfer matrices.

The glory of the corner transfer matrix is that, as opposed to the
usual transfer matrix, its eigenvalues can be determined exactly and
explicitly if the model is integrable \cite{Baxbook}. Its eigenvalues
have special analyticity properties in one of the couplings ($x$
here), which allows one to show that they must be independent of other
couplings. In the hard-square model, the eigenvalues of the corner
transfer matrix are the same along the entire first-order
line. Therefore, they can be computed by going to the far ``end'' of
this line, where in the Hamiltonian $|U/w|=|V/w|\to\infty$. In this
limit, not only are the three states $G_{12}$, $G_{23}$ and $G_{34}$
the exact ground states of the Hamiltonian, but every height
configuration is an eigenstate. This limit is ``frozen'', because in
the two-dimensional lattice model all the heights along a diagonal are
frozen to be the same. For example, in the northeast and southwest
quadrants of figure \ref{fig:CTM}, the heights along each diagonal in
the northwest/southeast direction must be the same.  Moving away from
the frozen limit, the eigenstates of the corner transfer matrix will
change, but the eigenvalues, and hence the type of ground states, stay
the same all along the first-order line. The same picture holds for
all $p$: there is a first-order transition line going from a
multicritical point to a frozen point.

{}From the corner transfer matrix eigenvalues one can compute the local height
probabilities $P(a|b,c)$, defined as the probability that the height
in the center will be height $a$, given that the heights around the
edges of the square are $b$ and $c$, as illustrated in figure \ref{fig:CTM}.
The $N$ heights along a line starting at the center
are labeled $h_i$, so that $h_1=a$, $h_{N-1}=b$, and
$h_{N}=c$. Then \cite{ABF}
\begin{equation}
P(a|b,c)= {\cal P}_a \sum_{\{h_2,h_3,\dots h_{N-2}\}}
  q^{E(h_1,h_2,\dots,h_{N})}
\label{Pq}
\end{equation}
\begin{equation}
E(h_1,h_2,\dots,h_{N}) = \sum_{j=1}^{N-2}j|h_j-h_{j+2}|
\label{Eh}
\end{equation}
where ${\cal P}_a$ is a (known) factor independent of the heights
other than $h_1=a$, and $q$ parametrizes the first-order line, so that
$q\to 0$ is the frozen limit and $q\to 1$ is the multicritical
point. The sum is over all allowed height configurations, and the
contribution of each distinct configuration to this probability is
obvious. The $j$ in the expression for $E$ arises because there are $j$ sites
along the $j$th diagonal, and in the frozen limit each site
contributes equally to the eigenvalue. 
  
The results found in the previous subsection \ref{TL-dilute} using the
Hamiltonian are therefore identical to those implied by the corner
transfer matrix, and generalize to all $p$. Ground states in the
frozen limit have $E=0$ in (\ref{Eh}). These are states where the
heights on every other site are the same, and it is natural to expect
that the ground states all along the first-order line will be
dominated by these configurations. Because of the restriction that
$|h_j-h_{j+1}|=1$, all states of the form $G_{j,j+1}\equiv (\dots
j,j+1,j,j+1\dots)$. For the $p$th model, this means there are $p-1$
ground states. A defect at site $j+1$ occurs when $|h_{j}-h_{j+2}|=2$;
such a defect runs all along the diagonal, and so has an energy of $2$
per unit length in the units implied by (\ref{Pq},\ref{Eh}). All
defects have this energy per unit length, no matter which ground
states they separate.

The translation of these results to the two-dimensional lattice model
is essentially the same as for $p=4$ \cite{Huse}. Each of the $p-1$
ground states $G_{j,j+ 1}$ corresponds to a degenerate minimum of
the free energy. Thus for $q\to 0$ the partition function is dominated
by regions of these minima, separated by domain walls which all have
the same energy per unit length. For general $p$, as opposed to
$p=3,4$, there is no symmetry forcing the domain walls to have the
same energy, but rather it is a consequence of the fine tunings
necessary to make the model integrable and on the first-order
transition line. As $q\to 1$, the multicritical point of interest is
approached, and the energy per unit length of the domain walls goes to
zero. 

\begin{figure}[h] 
\begin{center} 
\includegraphics[width= .49\textwidth]{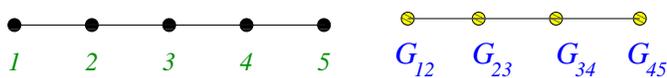} 
\caption{On the left, the adjacency diagram for the heights of the
  $p=5$ RSOS model; on the right, the adjacency diagram for the ground
  states of the same model.} 
\label{fig:adjacency} 
\end{center} 
\end{figure}

The restrictions on adjacent heights and on domain walls can be simply
encoded in terms of {\em adjacency diagrams}. For heights, the diagram
has a node for every allowed height, and two nodes are connected by a
edge if the corresponding heights are allowed to be on adjacent
sites. For ground states, the diagram has a node for every ground
state, and two nodes are connected if a fundamental domain wall can
separate the two ground states. 
The RSOS
models discussed above have
$p-1$ different ground states $G_{12}$, $G_{23}$, $G_{34},\dots$. 
The fundamental domain walls separate successive ground states on this
list: i.e.\ a region of ${G}_{h,h+1}$ can only be adjacent to 
$G_{h-1,h}$ and $G_{h+1,h+2}$. Thus in this case the adjacency
diagrams for the heights and for the ground states look the same,
except there are $p$ nodes in the height diagram, and $p-1$ nodes in
the ground-state diagram. For both types, the $j$th node is connected
to the $j+1$ and $j-1$ nodes. The nodes of the ground-state
diagram correspond to the edges of the height diagram; this remains
true for the more general RSOS models discussed in the next section.
The two diagrams for $p=5$ are displayed
in figure \ref{fig:adjacency}. 

By the same argument as for $p=4$, these results make it likely that the
$p$th minimal conformal field theory describes the continuum limit of
the critical point of a loop model with partition function of the form
(\ref{Zloop},\ref{Zloops}):
\begin{equation}
Z_{p}\approx \sum_{\cal L} w({\cal L}) e^{-f_D L}.
\label{Zloops2}
\end{equation}
with suitable boundary conditions. To obtain the topological weight
$w({\cal L})$, one must sum over the all the height configurations
consistent with a fixed loop configuration.

The simplest way to do this sum is to define the {\em adjacency
matrix} ${\cal A}$ for the ground states. This is the adjacency
diagram in matrix form: each row and column of this matrix corresponds
to a ground state, with the ${\cal A}_{ab}= 1$ if the ground states
labeled by $a$ and $b$ can be separated by a fundamental domain wall,
and zero otherwise. For the RSOS models here, it is
\begin{equation}
{\cal  A}_{rs}=\delta_{|r-s|,1}
\label{ARSOS}
\end{equation}
 for $r,s=1,\dots,p-1$. For 
example, for $p=5$ it is 
$${\cal A}=
\begin{pmatrix}
0&1&0&0\\
1&0&1&0\\
0&1&0&1\\
0&0&1&0
\end{pmatrix}
\ .$$
The topological weight $w({\cal L})$ is the number of height
configurations possible for each loop configuration ${\cal L}$. This is
easily written in terms of ${\cal A}$.  Index each loop by $j$, such
that it separates regions of ground states labeled by $a(j)$ and
$b(j)$. Then the number of allowed height configurations is
$$ \sum_{a(j),b(j)}
\prod_{j=1}^{\cal N} {\cal A}_{a(j),b(j)}$$
where the sum is over all ground states in each region. Thus adding
another loop amounts to multiplying by ${\cal A}$. For a large
number of non-intersecting loops ${\cal N}$, one therefore obtains
$$w({\cal L})\approx \lambda^{\cal N}$$
where $\lambda$ is the largest eigenvalue of ${\cal A}$. 
For the adjacency matrix (\ref{ARSOS}), this is
$$\lambda=2\cos\left(\frac{\pi}{p}\right).$$
Thus the RSOS models in this section indeed correspond in the
continuum limit to the $O(n)$ loop model with $n=2\cos(\pi/p)$. This
agrees with the two earlier cases we derived: $n=1$ for the Ising
model, and $n=\sqrt{2}$ for the hard-square model.

A very nice corroboration of this picture comes from studying the
scattering matrix of the 1+1 dimensional theory in the continuum limit
of the first-order line \cite{ZRSOS}. In field-theory language, moving
along the first-order line away from the multi-critical point
corresponds to a perturbation by the relevant $\Phi_{1,3}$ ``energy''
operator. This field theory is integrable like the underlying lattice
model, so one find the quasiparticles and their exact
scattering matrix. These quasiparticles turn out to be kinks which are
defects between $p-1$ ground states, exactly as we have seen. This
identification of the worldlines of these kinks as the loops in
the $O(n)$ model with $n=2\cos(\pi/p)$ was made long ago
\cite{Zpoly}. In fact, the correspondence is even closer: the
scattering matrix itself can be expressed in terms of the
Temperley-Lieb algebra \cite{Smirnov92}. Namely, there is a
representation of $I$ and $e$ acting on the kinks, such that the
two-particle scattering matrix for two particles of momentum $p_1$ and
$p_2$ is
\begin{equation}
S(p_1,p_2) = f(p_1,p_2) \left( I + g(p_1,p_2) e\right)
\label{Smat}
\end{equation}
where $f$ and $g$ are known functions. This correspondence between the
loops and the world-lines of the kinks in the 1+1-dimensional
description was exploited in [\onlinecite{Fendley05}] to discuss different
loop models for the $Q$-state Potts models.

\section{New fully packed loop models}
\label{sec:FPL}

In section \ref{On-cft} I gave a dilute and fully packed loop model
associated with each RSOS height model of Andrews, Baxter and
Forrester. The conformal minimal models describe the critical points. 
In this and the next section the methods developed in the last
section are used to find new loop models from more complicated RSOS
height models. It is then natural to conjecture that the conformal
field theories describing the critical points of these height models
also describe critical points in these loop models.

\subsection{Fused RSOS models}

The Boltzmann weights of the integrable lattice models studied here
satisfy the Yang-Baxter equation. This very strong constraint allows
many properties of the models to be computed exactly, like the
spectrum of the corner transfer matrix discussed above. 

Another thing
the Yang-Baxter equation allows is a way to construct new integrable
lattice models from known ones. This procedure is called {\em fusion},
and was invented in [\onlinecite{KRS}], and applied to height models in
\cite{DJMO}. The fused models obtained from the RSOS models are
labeled by an integer $k$, and
each model within a series is labeled by the same integer $p$ as before. 
Thanks to corner-transfer matrix computations \cite{DJKMO}, the
conformal field theories describing the critical points of the fused
RSOS models are known: they are the coset conformal field theories
(\ref{coset}), which have central charge (\ref{ccoset}).

The states of the general integrable RSOS models are easiest to
understand in terms of representations of a quantum-group algebra. A
quantum-group algebra $U_q(G)$ is a one-parameter deformation of a
simple Lie algebra $G$. For generic values of the parameter $q$,
$U_q(sl(2))$ has irreducible spin-$s$ representations corresponding to
those of ordinary $sl(2)$, but when $q$ is a root of unity such that
$q^{2(p+1)}=1$, only those with $s< p/2$ are irreducible.  The tensor
product of the spin-$s$ representation with the spin-$1/2$ one is
\begin{equation}
(s)\otimes (1/2) = (s+1/2) + (s-1/2)
\label{su2q}
\end{equation}
as long as all representations involved have $s<p/2$, so e.g.\
$((p-1)/2)\otimes (1/2) =(p/2-1)$. Except for this truncation, this is
the same rule as for ordinary $sl(2)$.

Each allowed height $h$ of a $U_q(sl(2))$ RSOS model corresponds to
the irreducible representation of spin $(h-1)/2$, so the heights run
from $1,\dots p$. Each fused model is labeled by an integer $k$.  In
the $k$th model, the height $h_1$ is allowed to be adjacent to a
height $h_2$ if the corresponding representations of $U_q(sl(2))$ have
the tensor-product decomposition
\begin{equation}
\left(\frac{h_1 -1}{2}\right) \otimes \left(\frac{k}{2}\right) =
\left(\frac{h_2-1}{2}\right) + \dots
\end{equation}
This can be interpreted physically by saying that the link variables of
the fused models have spin $k/2$, with the original RSOS models
discussed in section \ref{sec:TL-RSOS} having $k=1$.
For $k=1$, using the tensor product (\ref{su2q}) indeed gives 
the adjacency diagram illustrated in figure \ref{fig:adjacency}. 
The tensor products of higher-spin
representations of $U_q(sl(2))$ can be built up from (\ref{su2q}). 
The fused RSOS models to be
discussed in this section have $k=2$, so the links have spin $1$.  
The tensor product of a spin-1 representation  can be found by
using the fact that $(1/2)\otimes (1/2) =
(0) + (1)$. Thus
$$(s)\otimes(1) = (s)\otimes(1/2)\otimes(1/2)\ -\ (s).$$
For $s<p/2-1$, the tensor product of $(s)$ with $(1)$ is thus the same
as for ordinary $sl(2)$. 
The different rules come from the
restriction that $s<p/2$. One has $(p/2 -1) \otimes (1) = (p/2-1) +
(p/2-2)$, while $((p-1)/2) \otimes (1) = ((p-3)/2)$. Note that $((p-1)/2)$
does not appear on the right side of the latter, despite its being an
allowed representation. 

For the general model labeled by $k$ and $p$, continuing in this
fashion gives the constraints on
nearest-neighbor heights $h_i$ and $h_j$ to be \cite{DJMO}
\begin{eqnarray}
\nonumber
h_i &=& 1,2\dots p \\
\label{constraints}
h_i-h_j &=& -k,-k+2,\dots, k\\
\nonumber
k+1<h_i+h_j &<& 2p-k+1
\end{eqnarray}
These rules are
symmetric under the interchange $h\to p+1-h$. The Boltzmann weights
for $k=1$ satisfy this symmetry as well, so those for higher $k$ found
by fusion satisfy this as well.
For $k$ even, all heights in a given configuration must be even or
odd. It then is consistent to restrict them all to be even, because
one obtains the same field theory for even or odd.

\subsection{Fully packed loop models for $k=2$}
\label{FPLk2}

I show here how to derive fully packed loop models for the $k=2$ fused
RSOS models. This is done in a similar fashion as for the $k=1$ case,
by rewriting the transfer matrix in terms of the generators of an
algebra, and then finding a graphical representation
generalizing that of the Temperley-Lieb algebra.

The transfer matrix for all $k$ can be written in the form
(\ref{Tprod}). For $k=2$, the weights for each plaquette are then
\begin{equation}
{\cal T}_i = I_i + x X_i + y E_i
\label{TXE}
\end{equation}
where $x$ and $y$ are parameters related by
\begin{eqnarray*}
y&=&x + Q \frac{x^2}{1-x}\ ,\\
\sqrt{Q}&=& q+q^{-1}=2\cos(\pi/(p+1))\ .
\end{eqnarray*}
The model is isotropic when
$x=1/(\sqrt{Q}+1)$ and $y=1$. The Hamiltonian
\begin{equation}
H = -\sum_i (X_i + E_i)
\label{HXE}
\end{equation}
follows from the $x\to 0$ limit.  The representation of $X_i$ and
$E_i$ in terms of the heights is given in appendix \ref{app:XE}. There
is one important difference between $X_i$ and the Temperley-Lieb
generators: $X_i$ can be non-zero when both pairs of
heights across a plaquette are different (i.e.\ $h_i\ne h_i'$ and
$h_{i-1}\ne h_{i+1}$). This has important consequences for both the
dilute and fully packed loop models.

The lattice model with weights (\ref{TXE}) is integrable when the
$X_i$ and $E_i$ satisfy an algebra known as the $SO(3)$
Birman-Murakami-Wenzl (BMW) algebra \cite{BMW}. This algebra is best
thought of as the spin-1 generalization of the Temperley-Lieb
algebra. In fact, any representation of the Temperley-Lieb algebra
yields one of the $SO(3)$ BMW algebra \cite{fendleyread}; expressions
for $X_i$ and $E_i$ in terms of the $e_i$ are given in appendix
\ref{app:XE}.  The algebraic relations of the $X_i$ and $E_i$ can be
found in [\onlinecite{fendleyread}], and are given in appendix \ref{app:XE}.


As with lattice models based on the Temperley-Lieb algebra, many of
the properties of the models with Boltzmann weights (\ref{TXE}) are
independent of the representation of the $X_i$ and $E_i$. The most
illuminating representation of these generators is a graphical one.
To find this, it is useful to study first another representation of
the same algebra \cite{CZ,fendleyread}. 

This representation arises in the $Q$-state Potts model with
infinitely-strong antiferromagnetic nearest-neighbor interactions, and
ferromagnetic next-nearest-neighbor interactions. A ``spin'' $s_i$,
taking the integer values $s_i=1\dots Q$, is placed at each site $i$
of the square lattice. The restriction on nearest-neighbor spins is
simpler than that of the height models: adjacent spins must be
different. If all configurations were weighted equally, this would be the
antiferromagnetic $Q$-state Potts model at zero temperature. However,
at the critical point of interest, there are ferromagnetic
next-nearest-neighbor interactions. Labeling the spins around a
plaquette as $s_{i-1}$, $s_i$, $s_{i+1}$ and $s_i'$, in the same
fashion as the heights in figure \ref{fig:diag}, the restriction on
nearest neighbors means that
$$\Gamma_i \equiv \delta_{s_{i-1} s_{i}} \delta_{s_i s_{i+1}}
\delta_{s_{i+1}s_i'} =1$$ for each plaquette. The transfer matrix is
then given by (\ref{TXE}), with
\begin{eqnarray}
\nonumber
I_i&=& \Gamma_i  \delta_{s_i s_i'}\\
\label{XEPotts}
X_i& =& \Gamma_i \\
\nonumber
E_i&=& \Gamma_i \delta_{s_{i-1}s_{i+1}}
\end{eqnarray}
It is easy to check that these satisfy the $SO(3)$ BMW algebra with
$Q=(q+q^{-1})^2$ \cite{fendleyread}. For $Q\le 4$ there is a critical point
described by a
conformal field theory with $c=1$ for $Q=3$ and $c=3/2$ for $Q=4$. 
Using (\ref{XEPotts}) in the Hamiltonian (\ref{HXE}) makes it clear
that the next-nearest-neighbor interaction wants to make spins the
same, so this critical point presumably separates an ordered phase
from a disordered one.

\begin{figure}[h!] 
\begin{center} 
\includegraphics[width= .41\textwidth]{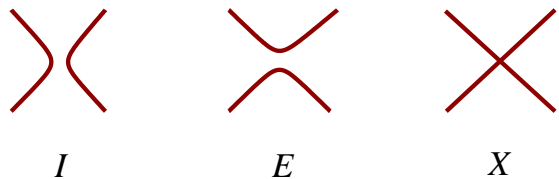} 
\caption{The graphical representation of the $SO(3)$ BMW algebra} 
\label{fig:IXE} 
\end{center} 
\end{figure}
A graphical representation of $I$, $E$ and $X$ in (\ref{XEPotts}) is
presented in figure \ref{fig:IXE}. The lines represent domain walls
between different spins; any spins not separated by a domain wall must
be the same. The $X$ generator has regions touching only at a point;
these may (but need not) have the same spin.

A fully packed loop model is obtained in the same fashion as the
Temperley-Lieb case. The partition function is expanded out in powers
of $x$ and $y$. Each term corresponds to a loop configuration after
the heights are summed over. The generators $I$ and $E$ are
90-degree rotations of each other, so a rotationally-invariant model
is obtained by taking $y=1$ and $x=1/(\sqrt{Q}+1)^{-1}$ in
(\ref{TXE}).  This yields a loop model with a typical
configuration illustrated in figure \ref{fig:FPL-BMW}.  The loops here
have a very important difference with the self- and mutually-avoiding
loops of the $O(n)$ loop model: because of the $X$ vertex, the loops
can touch. (Thus properly speaking, they shouldn't even be called
loops, but rather graphs with quadrivalent vertices.)

Since the $k=2$ RSOS height models are based on a different
representation of the same $SO(3)$ BMW algebra, these have the same
partition function as the loop models when
$\sqrt{Q}=q+q^{-1}=2\cos(\pi/(p+1))$ for $p$ integer. Therefore the
loop models at these values of $p$ also have critical points with the
same conformal field theory description, the superconformal minimal
models. The graphical representation is valid for all $Q$, not just
the $Q$ integer where (\ref{XEPotts}) applies, or for
$Q=4\cos^2(\pi/(p+1))$ with $p$ integer for the height models.  It is
likely that these loop models have a critical point for all $Q\le 4$,
although its field-theory description will only be unitary for $p$
integer.

The topological weight can be worked out using the BMW
algebra. However, it is much simpler to find it by using the
representation (\ref{XEPotts}) valid for integer $Q$. The lines in the
graphical representation correspond to domain walls for the spins
$s_i$.  This constrains spins in regions separated by domain walls
to be different (if two regions meet only at a point, the
corresponding spins are
still permitted to be the same). Each allowed spin configuration
then receives the same weight, so the topological weight per loop
configuration is given by the number of different spin
configurations. This is simply the number of ways of coloring the
regions with $Q$ colors, subject to the constraint that adjacent
regions have different colors. This is simply $\chi_Q$, as defined in
section \ref{sec:results}. This argument holds true for any positive
integer $Q$. This can be generalized uniquely to arbitrary $Q$ by
recognizing that $\chi_Q$ for a given dual graph satisfies the
recursion relation (\ref{recursion}), and thus is a
polynomial in $Q$. Thus the sum over heights for any $Q$ yields
$\chi_Q$. To get the full topological weight, note that every time an
$X_i$ appears in the expansion of the partition function, it comes
with a weight $x=1/(\sqrt{Q}+1)$, as detailed after equation
(\ref{TXE}). The number of $X$ vertices ${\cal N}_X$ is a topological
invariant.  The topological weight is therefore that given in
(\ref{wchrome}), namely
$$
w({\cal L})=\chi_Q({\cal L}) (\sqrt{Q}+1)^{{-\cal N}_X}\ .
$$

This  completes the mapping of the $k=2$ RSOS height models onto the
loop model described in section \ref{sec:results}. However, this 
fully packed loop model is not the only one in this universality class.
As described in appendix \ref{app:XE}, the
$SO(3)$ BMW algebra can be written in terms of the Temperley-Lieb
generators. In this representation, there are two lines on every
link, with a projection operator ${\cal P}_i$
ensure that the two are in the spin-1 representation of
$U_q(sl(2))$.  Therefore, the graphical representation in figure
\ref{fig:TL} can be used with (\ref{EXTL}) to give a fully packed loop
model on the $CuO_2$ lattice. The $CuO_2$ lattice is formed by taking
a square lattice (the coppers), and adding an extra site on each link
(the oxygens).  At every copper site, one gets $I$, $X$, or $E$ as
above. However, instead of using the above graphical representation
with an intersection for $X$, one writes them instead in terms of $I$
and $e$ using (\ref{EXTL}), and uses figure \ref{fig:TL} to represent them
graphically (illustrations can be found in [\onlinecite{fendleyread}]). The
projectors on the links (the oxygen sites) are expanded in terms of
$I$ and $e$ as well, by using ${\cal P}_i=1-e_i/(q+q^{-1})$. This means that on
the oxygen sites as well. This yields a somewhat strange but
well-defined fully packed loop model on the $CuO_2$ lattice, which has
a critical point described by the superconformal minimal models. This
construction can readily be generalized to larger $k$ integer as well.

\section{New dilute loop models}
\label{sec:dilute}

In this section I give a dilute-loop description of the $k=2$
height models and superconformal minimal models. The methods are
similar to those used in sections \ref{TL-dilute} and \ref{TL-CTM} for
the $k=1$ height models and the conformal minimal models. As opposed
to the precise results for the fully packed loop models, the mapping
for dilute loops is indirect, and thus is just a conjecture, not a proof.

The corner transfer matrix results for the fused RSOS models have long
been known \cite{DJKMO}. The results of interest are along an
integrable line of couplings, known as ``Regime III'' in the
literature. Like for the hard-square and $k=1$ RSOS models, along this
line, the Hamiltonian (\ref{HXE}) has degenerate ground states. This
is apparent from the computation of the local height probabilities
from the corner transfer matrix. The formula is identical to the
earlier case; the only distinction is that the restrictions on the
heights are different. Namely, the local height probability is still
given by (\ref{Pq}), where $q$ parametrizes the integrable line of
couplings, with $q\to 0$ the frozen limit and $q\to 1$ is the
multicritical point. 

\subsection{Dilute loop models for $k=2$}

\begin{figure}[h] 
\begin{center} 
\includegraphics[width= .4\textwidth]{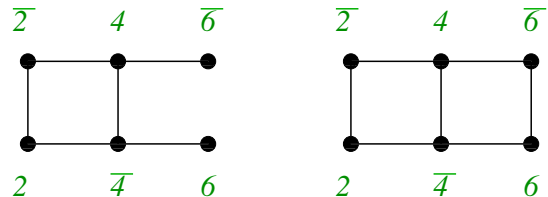} 
\caption{the adjacency diagrams for the heights of the
  fused RSOS model with $k=2$, $p=6$ (left) and $k=2$, $p=7$ (right).} 
\label{fig:adjBMW} 
\end{center} 
\end{figure}

For $k=2$, the heights are even integers $h_i \le
p$, and $h_i$ and $h_j$ on adjacent sites obey $h_i-h_j=0,\pm 2$ and
$2<h_i+h_j<2p$. It is convenient to distinguish between heights on the two
sublattices of the square lattice, each sublattice comprised of every
other site. Denoting heights on one sublattice without a bar, and one
with a bar, the adjacency diagrams for the $k=2$ height models with
$p=6$ and $p=7$ are given in figure \ref{fig:adjBMW}. Unbarred heights
are always adjacent to barred heights, and vice versa.

\begin{figure}[h] 
\begin{center} 
\includegraphics[width= .48\textwidth]{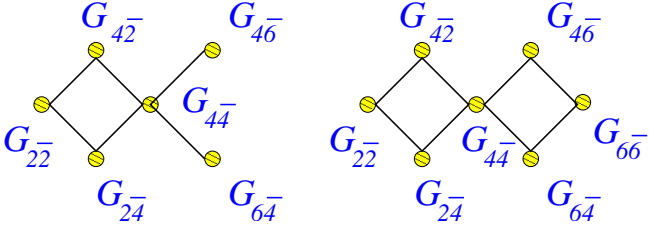} 
\caption{The ground-state adjacency diagrams for $k=2,p=6$ (left), and
  $k=2,p=7$ (right).}
\label{fig:adjgsBMW} 
\end{center} 
\end{figure}

The energy $E$ in (\ref{Eh}) is minimized when the heights on every
other site are the same. This is an exact ground state of the model in
the frozen limit. As argued in section \ref{TL-CTM}, each
configuration with this property should correspond to a ground state
of the Hamiltonian all along the integrable line, including at the
multicritical point, where $H$ is given in (\ref{HXE}). These ground
states are therefore of the form $G_{h,\bar{h}}=(\dots
h\bar{h}h\bar{h}\dots)$ and $G_{h,\overline{h\pm 2}}= (\dots h,\overline{h\pm
2},h,\overline{h\pm 2},\dots)$.  As before, each ground state corresponds
to a edge on the height adjacency diagram.  Excited states are defects
between ground states. For the $k=1$ case, a defect at site $i$ means
$h_{i+1}-h_{i-1}=\pm 2$.  Here, a defect can have $h_{i+1}-h_{i-1}=\pm
2$ or $\pm 4$.  However, for the latter defect, $E$ in (\ref{Eh}) is
twice as large. It is therefore natural to assume that $\pm 4$ defects
can be treated as two of the $\pm 2$ defects. (Thus one can have a
``double defect'' on one plaquette.) Thus fundamental defects
are only those where $h_{i-1}-h_{i+1}=\pm 2$, so the adjacency diagram
for ground states looks like those illustrated in figure
\ref{fig:adjgsBMW}. Another way of making the distinction between
fundamental and composite defects is that a fundamental defect occurs
between $G_{h_1\bar{j}_1}$ and $G_{h_2\bar{j}_2}$ if either $h_1=h_2$
or $j_1=j_2$.

The ground states and fundamental defects are easiest to list in terms
of new labels.  Relabel ground states of the form $G_{h,\bar{h}}$ as
${\cal G}^2_{h-1}$, those of the form $G_{h,\overline{h+2}}$ as ${\cal
G}^1_{h}$, and those of the form $G_{h+2,\bar{h}}$ as ${\cal
G}^3_{h}$. The different ground states ${\cal G}^r_s$ then have
$r=1\dots 3$, and $s=1\dots p-2$ with $r+s$ odd. The fundamental
defects then separate the ground states ${\cal G}^{r_1}_{s_1}$ and
${\cal G}^{r_2}_{s_2}$ if $r_1=r_2\pm 1$ {\em and} $s_1=s_2\pm
1$. Note that with this relabeling, the constraints on $r$ and $s$ are
independent of each other. Moreover, the constraints on $r$ and $s$
individually are exactly the same as for the ground states of the
$k=1$ height models. This means that the 
ground-state adjacency matrix for the $k=2$ models can be written as
the tensor product
\begin{equation}
{\cal A}^{2,p}={\cal A}^{1,4}\otimes{\cal A}^{1,p-1}
\label{A2p}
\end{equation}
where we denote the adjacency matrices for general $k$ and $p$ by
${\cal A}^{k,p}$. The adjacency matrix for the heights does {\em not}
obey such a simple decomposition: this is only a property of the
ground states and fundamental defects.

As with the $k=1$ case, the world lines of the fundamental defects are
interpreted in the two-dimensional lattice model as domain walls
between free-energy minima. These domain walls form the loops.
Domain walls in these RSOS models occur when heights {\em across} a
square plaquette are different. These domain walls are drawn on the
{\em diagonals} of the original lattice, as seen in the
(non-intersecting) loops of figure \ref{fig:hardsq}. (In the
fully packed loop models, the links are drawn on the links of the dual
lattice.)  Each link of the original lattice corresponds to a
ground state, so the domain walls separate links with different ground
states.

To complete the argument, the topological weight $w({\cal L})$ of each
loop configuration must be found. The simplest possibility, realized
for $k=1$, is that is that these domain walls never cross, so that
the loops self-avoid and mutually avoid. Summing over the heights in
the model then gives the topological weight of the $O(n)$ loop model:
$w=n^{\cal N}$, where ${\cal N}$ is the number of loops. For several
reasons, such a $w$ cannot be correct for $k>1$.

First of all, domain walls do cross for $k>1$. This happens when
all four ground states around a
link are different. In the $k=2$ model, the domain walls cross
when the heights around a plaquette are of the forms
\begin{equation}
\begin{picture}(200,55)
\put(28.5,0){$h$} \put(7,23){$\overline{h}$} \put(30,10){\line(-1,1){15}}
\put(30,10){\line(1,1){15}} \put(51,23){$\overline{h+2}$}
\put(45,25){\line(-1,1){15}} \put(15,25){\line(1,1){15}}
\put(20.5,45){${h+2}$} \put(90,22){\hbox{}}
\put(140.5,0){$h+2$} \put(127,23){$\overline{h}$} \put(150,10){\line(-1,1){15}}
\put(150,10){\line(1,1){15}} \put(171,23){$\overline{h+2}$}
\put(165,25){\line(-1,1){15}} \put(135,25){\line(1,1){15}}
\put(148.5,45){$h$} 
\end{picture}
\label{plaquettes}
\end{equation}
The four different ground states on these two
plaquettes are $G_{h,\overline{h}}$ , 
$G_{h+2,\overline{h}}$ , 
$G_{h+2,\overline{h+2}}$ and $G_{h,\overline{h+2}}$ .  These have a
non-zero Boltzmann weight coming from $X_i$, as given in appendix
\ref{app:XE}.  Configurations like these cannot occur in the $k=1$
model, where $I_i$ requires that the top and bottom heights be the
same, while $e_i$ requires that the left and right be the same. 

Even if one were to assume that such crossings are irrelevant, the value of $n$ obtained is still inconsistent. Namely,
if crossings are ignored, one obtains the dilute $O(n)$ loop
models as with $k=1$. The arguments of section \ref{TL-CTM} give the
weight $n$ per loop to be the largest eigenvalue of ${\cal A}$. This
is easy to find from (\ref{A2p}): it is simply the product of the
largest eigenvalues of ${\cal A}^{1,4}$ and ${\cal A}^{1,p-1}$, which
are $2\cos(\pi/4)=\sqrt{2}$ and $2\cos(\pi/(p-1))$ respectively. This
would yield $n=2\sqrt{2}\cos(\pi/(p-1))$. Although this is the precise
value of $n$ implied by the SLE results \cite{Gruzberg05}, the loop
model cannot be the $O(n)$ model. The weight per loop has $n>2$ (e.g.\
$n=2\sqrt{2}$ for the $SU(2)_2$ WZW model) in general, and the $O(n)$
model in this regime does not have a critical point. Since the
underlying lattice model is critical, a non-critical loop model
obviously cannot be an equivalent description. The fact that domain
walls can cross or touch must affect the continuum behavior of the
loop model.

The fact that the weight of non-crossing loops is the same as the SLE
results is quite encouraging. The task is then to use the Boltzmann
weights of the height model to understand how to treat the crossings.
The fully packed loops do intersect, and affect the topological weight
via the chromatic polynomial. The dilute loops behave differently. The
decomposition (\ref{A2p}) suggests that for $k>1$ the domain walls
should not be treated as a single lines, but instead split into {\em
two} lines, {\em each} of which behaves as a $k=1$ model.

\begin{figure}[h] 
\includegraphics[width= .48\textwidth]{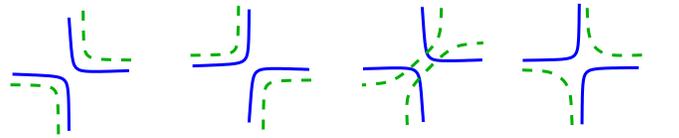} 
\caption{The four ways of resolving a domain-wall crossing with
  doubled lines}
\label{fig:dilute-resolve} 
\end{figure}

The key to defining a topological weight is the relabeling of the
ground states as ${\cal G}_r^s$ as described above. In this
relabeling, the $r$ and $s$ labels behave independently, as far as the
adjacency restrictions on the ground states are concerned. Since the
domain walls are between the ground states, this means the domain
walls in the $r$ and $s$ labels can be treated independently. Thus the
domain walls can be split into a ``solid'' wall and a ``dashed'' wall,
as illustrated in figure \ref{fig:dilute-k}. Away from where the
domain walls meet, this splitting is unimportant, and the solid and
dashed lines are attached.  However, at a domain-wall crossing there
are now {\em four} ways of resolving the lines so that lines of the
same type do not cross. These are illustrated in figure
\ref{fig:dilute-resolve}.
The dashed line is a domain wall in the
$1,4$ theory (the $r$ index in ${\cal G}_s^r$), while the solid line
corresponds to a domain wall in the $1,p-1$ model (the $s$ index in
${\cal G}_s^r$).

Note that dashed lines never cross or touch each other, and likewise
for solid, but that the dashed and solid can cross one another. Thus a
loop configuration consists of dashed and solid loops which are sewn
together, except when two loops meet on one plaquette. At each
domain-wall crossing, one of the four possibilities in figure
\ref{fig:dilute-resolve} must occur. A typical loop configuration is
illustrated in figure \ref{fig:dilute-k}.  Since the restrictions on
the $r$ and $s$ labels are completely independent, it is consistent to
treat the dashed and solid domain walls independently.

All the configurations in the height model 
can therefore be realized in terms of a pair of $k=1$ domain walls.
The last two domain-wall configurations in figure \ref{fig:dilute-resolve}
are necessary to realize configurations like (\ref{plaquettes}) that occur 
for $k>1$.
This follows
from rewriting the ground states around the plaquette
in terms of ${\cal G}^r_s$. For example,
$G_{h,\overline{h}}={\cal G}^2_{h-1}$ and  
$G_{h+2,\overline{h+2}}={\cal G}^2_{h+1}$. Therefore they have the
same $r$ value and are not separated by any dashed lines. Likewise
$G_{h+2,\overline{h}}={\cal G}^3_h$ and $G_{h,\overline{h+2}}={\cal G}^1_h$ 
so these have the same $s$ value and are not separated by any solid
lines. 

This picture is in remarkable accord with the field-theory results for
the ($k=2$) superconformal minimal models perturbed along the
integrable line of interest.  The particles in the field theory 
have the same labels as
the defects, so the particle world-lines can again be identified
with the domain walls. Moreover, the scattering matrix $S^{2,p}$ is a
product of $k=1$ scattering matrices \cite{Schoutens90}, just like the
ground-state adjacency matrices. It is
\begin{equation}
S^{2,p}= S^{1,4} \otimes S^{1,p-1}.
\label{S2p}
\end{equation}
Plugging the $k=1$ scattering matrices from (\ref{Smat}) into
(\ref{S2p}) shows that $S^{2,p}$ is a sum of four terms, involving
$I\otimes I$, $e\otimes I$, $I\otimes e$, and $e\otimes e$. Recalling
the graphical representation of $I$ and $e$ in figure \ref{fig:TL}
gives precisely the four possibilities in figure
\ref{fig:dilute-resolve} for the particle world-lines.

It is now clear how to build a topological weight for these dilute
loops. The dashed and solid lines each form closed loops. The sum over
heights can be done independently for the $r$ and $s$ labels for each
ground state.  The weight per dashed loop is then simply the largest
eigenvalue of the ${\cal A}^{1,4}$ adjacency matrix, which is
$\sqrt{2}$. The weight per solid loop is the largest eigenvalue of
${\cal A}^{1,p-1}$, which is $2\cos(\pi/(p-1))$.  When the numbers of
dashed and solid loops are ${\cal M}$ and ${\cal N}$ respectively, the
topological weight is
\begin{equation}
w({\cal L}) = \sqrt{2}^{\cal M}
\left(2\cos\left(\frac{\pi}{p-1}\right)\right)^{\cal N} b^{\cal C}\ .
\end{equation}
The topological invariant ${\cal C}$ is the number of times the
(original) domain walls cross, i.e. the number of plaquettes which
have one of the configurations in figure \ref{fig:dilute-resolve} at their
center. The piece involving the parameter $b$ 
is analogous to the last piece of the fully packed weight
(\ref{wchrome}). The arguments here are not refined enough to
determine the value of $b$, or whether changing it results in a
relevant or irrelevant perturbation of the critical point.

\subsection{Arbitrary $k$}

The ground states for general $k$ are quite analogous to those for
$k=1$ and $k=2$. The expression (\ref{Pq},\ref{Eh}) for the local
height probabilities still holds, so the minimum of $E$ occurs when
the heights on every other site are the same. Thus each ground state
can be labeled by two heights $h_i$ and $h_j$, which must satisfy the
constraints for nearest-neighbor heights in (\ref{constraints}). 

For general $k$ and $p$ integer, the adjacency matrix
decomposes into tensor products of $k=1$ adjacency matrix. This
follows by relabeling the ground states $G_{h_1,h_2}$
into the form ${\cal G}^r_s$ via
\begin{eqnarray*}
r&=& \frac{h_1-h_2+k+2}{2}\ ,\\
s&=& \frac{h_1+h_2-k}{2}\ .
\end{eqnarray*}
The constraints in (\ref{constraints}) in terms of $r$ and $s$ are
therefore quite simply
$r=1,2\dots k+1$
and $s=1,2,\dots p-k$.

A fundamental defect needs to have energy $E=2$
in ($\ref{Eh}$). This only occurs if the two ground states have one of
their height labels in common, and one different. Namely, a
fundamental defect separates the ground state
$G_{h_1,h_2}$ from the ground states $G_{h_1,h_2\pm 2}$ and $G_{h_1\pm
  2,h_2}$. In terms of $r$ and $s$ labels, this means that
a fundamental defect separates ${\cal G}^r_s$ from the states ${\cal G}^{r\pm
  1}_{s\pm 1}$ and ${\cal G}^{r\mp   1}_{s\pm 1}$. Thus the
restrictions on $r$ and $s$ are independent of each other, and the
adjacency matrix for the ground states for general $k$ and $p$ is simply
\begin{equation}
{\cal A}^{k,p}={\cal A}^{1,k+2}\otimes{\cal A}^{1,p-k+1}
\end{equation}
This is in harmony with the scattering matrix for the quasiparticles
in the field theory describing the integrable line, which is
\cite{Zamolodchikov:1991vg}
\begin{equation}
S^{k,p}=S^{1,k+2}\otimes S^{1,p-k+1}.
\end{equation}
This also is in harmony with the Coulomb-gas approach to these
conformal field theories, which gives their partition functions as
sums of products of minimal-model ones \cite{DSZ88}.

The configurations in the dilute loop model for general $k$ are
therefore the same as for $k=2$ model.
The defect world-lines become ``double''
domain walls in the two-dimensional picture. Crossings are
resolved in the four ways in figure (\ref{fig:dilute-resolve}), so the typical
configuration displayed in figure \ref{fig:dilute-k} still is
applicable here. The weight for the dashed loops for general $k$ is now
the largest eigenvalue of ${\cal A}^{1,k+2}$, which is
$2\cos(\pi/(k+2))$. This yields the topological weights given in
(\ref{wnm},\ref{nm}).

\section{Further directions}

In this paper I have shown how to turn lattice models into loop
models. Dense loop models are found directly by exploiting the
algebraic structure of the transfer matrix, while dilute loops are the
domain walls found more indirectly using the corner
transfer matrix and the scattering matrix. Since these lattice models
have critical points described by known coset conformal field
theories, this means that the loop models have these same critical
points as well. 

It would be interesting to show directly that the dilute loop models
have these critical points. One way of doing so would be to study the
``dilute BMW models'' \cite{Grimm}. These lattice models generalize
the dilute Temperley-Lieb models \cite{Nienhuis90,Warnaar93a}
discussed above to allow a more general algebra. The domain walls for
the dilute $SO(4)$ BMW model are exactly the same doubled domain walls as in
the dilute loop model discussed here (because the algebra
$SO(4)=SU(2)\times SU(2)$). The corner transfer matrix computation for 
these models has not yet been done, so it is not yet known if these
models have the critical points described in this paper.

It would also of course be interesting to generalize these results to
other loop models and conformal field theories. Another loop model
with a critical point described by the conformal minimal models was
discussed at length in [\onlinecite{Fendley05}]. This dilute loop model can be
derived from the Potts lattice model from both the domain-wall and the
algebraic approach (using projection operators instead of $I,X$ and
$E$), but the heuristic approach used there is more suggestive. This
is the scattering matrix approach used above, where the world lines of
the particles in the field theory are identified with the loops. As
mentioned briefly in [\onlinecite{Fendley05}], the scattering matrix
approach suggests that the dilute version of the
fully packed $k=2$ model discussed above will have critical points
described by parafermion conformal field theories. This could be
tested directly by studying the $SO(3)$ dilute BMW model of
[\onlinecite{Grimm}].

Many integrable lattice models and scattering matrices are known, and
it likely that the methods described here could be
applied to them. Finding more examples would be of great interest in both
the areas mentioned in the introduction: SLE and non-abelian
statistics. Much is known about the relations between conformal field
theory, non-abelian statistics, topology and geometry in the simplest
cases. Hopefully the generalizations described here will be of use in
deepening this knowledge.

\bigskip

\noindent
{\bf Acknowledgments}
I am grateful to John Cardy, Eduardo Fradkin, Michael
 Freedman, Ilya Gruzberg, Andreas Ludwig, Nick Read, Kevin Walker, and Paul
 Wiegmann for many interesting conversations on loop models and related topics.
This research has been supported by the NSF under grant
DMR-0412956.

\appendix

\section{A brief review of the conformal field theories}
\label{app:CFT}

Field theories describing rotationally-symmetric two-dimensional
classical critical points are invariant under conformal
transformations of two dimensional space.  Conformal symmetry in two
dimensions is very powerful, because it has an infinite number of
generators: representations of the corresponding symmetry algebra
(called the Virasoro algebra) are infinite-dimensional.  {\em Minimal
models} are conformal field theories which have a finite number of
highest-weight states under the conformal symmetry \cite{Belavin84}.
These are among the models for which SLE is applicable. 
The minimal models have been classified, and all the critical exponents are
known \cite{Belavin84,FQS}. Many (and in principle, all) of the
correlation functions can be explicitly computed.
A convenient way of labeling conformal field theories is in terms of a
number $c$ called the {\em central charge}. Unitary minimal models have
central charge
\begin{equation}
c=1-\frac{6}{p(p+1)}\ .
\label{cmin}
\end{equation}
where $p$ is an integer obeying $p\ge 3$.  These are the only unitary
conformal field theories with $c<1$ \cite{FQS}. Any theory with $c\ge
1$ has an infinite number of highest-weight states \cite{Cardy86}.

To study more general conformal field theories, it is useful
to extend the Virasoro algebra by some other generators. One
well-studied way of doing so is to extend the symmetry by a simple Lie
algebra $G$. These conformal field theories are called
Wess-Zumino-Witten (WZW) models \cite{Witten84}, and the
extended symmetry algebra is called a Kac-Moody algebra. Like with
the minimal models, all critical exponents are known, and correlators
can be computed explicitly \cite{KZ}. Each unitary WZW model $G_k$ is
labeled by the algebra $G$ and an integer $k\ge 1$, called the level.

Minimal models can be constructed from WZW models by using the {\em
coset} construction \cite{GKO}. A $G/H$ coset conformal field theory is defined
with energy-momentum tensor $T_{G/H}=T_G-T_H$, where $T_G$ and $T_H$
are the energy-momentum tensors of the $G$ and $H$ WZW models. This
allows many of the properties (e.g.\ the critical exponents) of the
coset models to be computed by using results from the WZW models. 
The coset models of interest here are
\begin{equation}
\frac{SU(2)_k \times SU(2)_{p-k-1}}{SU(2)_{p-1}}
\label{coset}
\end{equation}
with $p>k+1$. 
They have central charge
\begin{equation}
c=\frac{3k}{k+2}\left(1-\frac{2(k+2)}{(p+1)(p-k+1)}\right).
\label{ccoset}
\end{equation}
When $k=1$, these are the minimal models. The models with $k=2$ are
usually known as the ${\cal N}=(1,1)$ superconformal minimal models,
because they have an extended symmetry algebra including supersymmetry.
Note that the simplest superconformal
minimal model (with $k=2,p=4$)
is identical to the second conformal minimal model (with $k=1$,
$p=4$).  As $p\to\infty$ for
fixed $k$, one obtains simply the $SU(2)_k$ WZW model. The Coulomb-gas
formulation of these coset models was developed in [\onlinecite{DSZ88}].

In the loop models discussed below, $p$ becomes a continuous
parameter. The conformal field theories describing loop models for $p$
non-integer generally are expected to be non-unitary and not
rational. Exact computations are still possible by using the
Coulomb-gas technique within conformal field theory, and (for $c\le
1$) SLE.

\section{Loops on the sphere}
\label{app:bc}
All the results of this paper apply when two-dimensional space is
topologically a sphere. In this appendix I show how to implement these
boundary conditions with the transfer matrix written in terms of
Temperley-Lieb generators, and show that the partition functions in
this situation are independent of representation of this algebra.

In a loop
model, space is topologically a sphere when no loops end on the boundary. To
implement these boundary conditions in the transfer-matrix
formulation of the height model, define the operator 
$${\cal B}=(q+q^{-1})^N e_1 e_3\ \dots\ e_{2N-1},$$
which acts on a zig-zag row of heights as illustrated in figure
\ref{fig:diag}, acting on a set of heights $h_0,h_1,\dots
h_{2N}$. ${\cal B}$ is normalized so that ${\cal B}^2=\cal{B}$. 
Then the partition function for $M+1$ zig-zag rows is
$$Z_{height}= ({\cal B}T^M {\cal B})_{\{a\}\{a\}} $$ where
${\{a\}\{a\}}$ means the matrix element which has all the heights in
the first and last rows fixed to be $a,a+1,a,a+1,\dots$.  The fact
that $T$ involves only $I_0$ and $I_{2N}$ for the left- and right-most
columns means that all the heights along the sides are fixed to the
same values.

To relate this definition of $Z$ with that coming from the loops,
expand $T$ in terms of the Temperley-Lieb generators, as done for the
fully packed loop models. Each of the resulting terms corresponds to a
loop configuration ${\cal L}$, and consists of many $e_i$ sandwiched
in between the ${\cal B}$s.  Consider one of these terms, which gets
weight $A({\cal L})$ in the loop model, and $A_{\{a\}\{a\}}$ in the
height model.  Then using the algebra (\ref{TLalg}), I show here that
\begin{equation}
A({\cal L}) {\cal B}_{\{a\}\{a\}}= A_{\{a\}\{a\}}\ .
\label{ABA}
\end{equation}
When computing ${\cal A}({\cal L})$ the $e_i$ in ${\cal B}$ are drawn
with only one loop: the piece that goes ``off the edge'' is ignored.
The relation (\ref{ABA}) shows that up to an overall constant, 
the partition functions are the same:
\begin{equation}
Z_{\hbox{loop}} = {\cal B}_{\{a\}\{a\}} Z_{\hbox{height}}
\label{ZBZ}
\end{equation} 

Let $j$ be an integer. 
Because $(e_i)^2=(q+q^{-1})e_i$ and all $e_{2j-1}$ commute with each other,
$$
e_{2j-1}{\cal B}={\cal B}e_{2j-1}=(q+q^{-1}){\cal B}.
$$ Thus any time an $e_{2j-1}$ is next to ${\cal B}$, it can be
removed, leaving an overall factor $q+q^{-1}$. Now consider an $e_2$
somewhere in the middle. All one needs to do to get rid of it is 
to get the $e_1$s from the ${\cal B}s$ next
to the $e_2$, and then use $e_1 e_2 e_1 =
e_1$. Since $e_1$ commutes with everything other than $e_2$,
the only obstruction to doing this is the product is of
the form
$$e_1 e_2 \dots e_3 \dots e_2 e_1.$$
where $e_1$ does not appear in the dots. 
Since there are no $e_1$s
in the dots, $e_2$ can be commuted through the dots and 
$e_2 e_3 e_2 =e_2$ used, unless there is an $e_4$ within the
dots, i.e.\ the product is of the form
$$e_1 e_2 e_3 \dots e_4 \dots e_3 e_2 e_1.$$ The offending $e_4$ can
then be eliminated using the $e_3$s unless there is an $e_5$ which
interferes. Thus either the offending ones can be eliminated, or one
ends up with
$$e_1 e_2 e_3 e_4 \dots e_{2N-2} e_{2N-1} e_{2N-2}\dots  e_3 e_2
e_1.
$$
But then this can be reduced by using  $e_{2N-2} e_{2N-1}
e_{2N-2}=e_{2N-2}$ and so on. 
Proceeding like this one can eliminate all of them, leaving
\begin{equation}
A= (q+q^{-1})^{\cal N} {\cal B}
\end{equation}
where ${\cal N}$ is an integer independent of representation. In fact,
${\cal N}$ is the number of closed loops in the graphical
representation of ${\cal L}$.  This proves (\ref{ABA}) and (\ref{ZBZ})
for any representation of the $e_i$, as long as it satisfies the
Temperley-Lieb algebra.

For the loop models based on the $SO(3)$ BMW algebra, the $e$
in ${\cal B}$ are replaced with $E$. The arguments then go through as above.

\section{$E$, $X$ and BMW}
\label{app:XE}

This appendix collects formulas connected to the $SO(3)$ BMW
algebra. Much of this appendix is discussed in depth in
[\onlinecite{fendleyread}], and what is not can be obtained by using
these results along with the explicit Boltzmann weights for the $k=2$
RSOS models \cite{DJMO}.

The generators $X_i$
and $E_i$ can be defined in terms of the Temperley-Lieb generators
by using the projectors ${\cal P}_i=I-e_i/(q+q^{-1})$ onto 
the spin-1 representation of $U_q(sl(2))$, giving
\cite{fendleyread} 
\begin{eqnarray}
E_{i} &=& {\cal P}_{2i+1}\, {\cal P}_{2i+3}\, e_{2i+2}\,
e_{2i+1}\, e_{2i+3}\, e_{2i+2}\, {\cal P}_{2i+1}\, {\cal
P}_{2i+3},\cr\cr X_{i} &=& (q+q^{-1}){\cal P}_{2i+1}\, {\cal P}_{2i+3}\,
e_{2i+2}\, {\cal P}_{2i+1}\, {\cal P}_{2i+3}. \label{EXTL}
\end{eqnarray}
Representations of the $SO(3)$ BMW algebra therefore can be found from
representation of the Temperley-Lieb algebra simply by plugging into
(\ref{EXTL}). In particular, a graphical representation in terms of
non-crossing loops can be found from the pictures in figure
\ref{fig:TL} \cite{fendleyread}.

The $SO(3)$ BMW algebra is usually written in terms of $E_i$ and
$B_i$, where $B_i$ is a braiding operator (i.e.\ satisfies
$B_iB_{i+1}B_i=B_{i+1}B_iB_{i+1}$). Here it is convenient to write them
in terms of $I,X$, and $E$; the expression for $B$ in terms of these
can be found in [\onlinecite{fendleyread}]. The $E_i$ 
satisfy the Temperley-Lieb algebra amongst themselves:
\begin{eqnarray}
E_i^2 &=& (Q-1) E_i,\nonumber\\ E_i\, E_{i\pm 1}\, E_i &=&
E_i.
\label{TLalgE}
\end{eqnarray}
Relations involving the $X$ are:
\begin{eqnarray}
(X_i)^2 &=& (Q-2) X_i +  E_i.
\label{EXalg}
\end{eqnarray}
and 
\begin{eqnarray*}
X_iE_{i+1}X_i &=&
X_{i+1}E_iX_{i+1},\\ E_iX_{i+1}E_i  &=& (Q-1)E_i,\\
X_iE_{i+1}E_i  &=& X_{i+1}E_i,\\ X_iX_{i+1}E_i  &=&
(Q-2)X_{i+1}E_i + E_i,\\ X_iX_{i+1}X_i -  X_{i+1}X_iX_{i+1}
&=& X_{i+1}E_i + E_iX_{i+1} + X_i \\&&- E_i 
-X_iE_{i+1} - E_{i+1} X_i \\&&-  E_{i+1} +X_{i+1}.
\end{eqnarray*}
All relations also hold with $i$ and $i+1$ interchanged, and with the
order of products in each term reversed. All generators labeled by
sites $i$ and $j$ commute when $|i-j|>1$. It is instructive to use the
graphical representation in figure \ref{fig:IXE} to verify these
relations.

Explicit expressions for $E$ and $X$ in the $k=2$ height
representation discussed in this paper can be found either by
substituting (\ref{eRSOS}) into (\ref{EXTL}), or from the
explicit Boltzmann weights. Since the
$E_i$ satisfy the Temperley-Lieb algebra, they end up satisfying the
identical relation as the $e_i$, the only difference being that the
heights here are always even, with the adjacency rules in
(\ref{constraints}) for $k=2$. Define $[a]= \sin(a\pi/(p+1)).$ The
$X_i$ in the height representation are
$$
\begin{picture}(200,45)
\put(3,23){$h$} \put(28.5,0){$h$} 
\put(51,23){$h+2$} \put(20.5,45){$h+2$} 
\put(80,23){=\quad \Large $\frac{\sqrt{[h-1][h+3]}}{[h+1]}$}
\put(30,10){\line(1,1){15}} 
\put(30,10){\line(-1,1){15}}
\put(45,25){\line(-1,1){15}} \put(15,25){\line(1,1){15}}
\end{picture}
$$
$$
\begin{picture}(200,45)
\put(3,23){$h$} \put(20.5,0){$h+2$} 
\put(51,23){$h+2$} \put(20.5,45){$h+2$} 
\put(80,23){=\quad \Large $\frac{[h+3]}{[h+1]}$}
\put(30,10){\line(1,1){15}} 
\put(30,10){\line(-1,1){15}}
\put(45,25){\line(-1,1){15}} \put(15,25){\line(1,1){15}}
\end{picture}
$$
$$
\begin{picture}(200,45)
\put(3,23){$h$} \put(20.5,0){$h+2$} 
\put(51,23){$h$} \put(20.5,45){$h+2$} 
\put(80,23){=\quad \Large $\frac{[h+2][2]}{[h+1][1]}$}
\put(30,10){\line(1,1){15}} 
\put(30,10){\line(-1,1){15}}
\put(45,25){\line(-1,1){15}} \put(15,25){\line(1,1){15}}
\end{picture}
$$
$$
\begin{picture}(200,45)
\put(3,23){$h$} \put(20.5,0){$h+2$} 
\put(51,23){$h$} \put(28.5,45){$h$} 
\put(80,23){=\quad \Large $\frac{[h-1]}{[h+1]}\sqrt{\frac{[h+2]}{[h]}}$}
\put(30,10){\line(1,1){15}} 
\put(30,10){\line(-1,1){15}}
\put(45,25){\line(-1,1){15}} \put(15,25){\line(1,1){15}}
\end{picture}
$$
$$
\begin{picture}(200,55)
\put(3,23){$h$} \put(28.5,0){$h$} 
\put(51,23){$h$} \put(28.5,45){$h$} 
\put(80,23){=\quad \Large $\frac{[1]}{[2][h]}
\left(\frac{[h-1]^2}{[h+1]} + \frac{[h+1]^2}{[h-1]}\right)$}
\put(30,10){\line(1,1){15}} 
\put(30,10){\line(-1,1){15}}
\put(45,25){\line(-1,1){15}} \put(15,25){\line(1,1){15}}
\end{picture}
$$ The remaining non-zero elements of $X_i$ are found by either using
the symmetry $h\leftrightarrow p+1-h$, or by flipping them
horizontally or vertically. All others are zero. The fact that the
first of these is non-vanishing has a number of interesting
consequences discussed in section \ref{sec:dilute}.


\def\prl#1#2#3{Phys.\ Rev.\ Lett.\ {\bf #1}, #2 (#3)}
\def\pra#1#2#3{Phys.\ Rev.\ A {\bf #1}, #2 (#3)}
\def\prb#1#2#3{Phys.\ Rev.\ B {\bf #1}, #2 (#3)}
\def\prbrc#1#2#3{Phys.\ Rev.\ B {\bf #1} [RC], #2 (#3)}
\def\prd#1#2#3{Phys.\ Rev.\ D {\bf #1}, #2 (#3)}
\def\pre#1#2#3{Phys.\ Rev.\ E {\bf #1}, #2 (#3)}
\def\physrev#1#2#3{Phys. Rev. {\bf #1}, #2 (#3)}
\def\npb#1#2#3{Nucl.\ Phys.\ B {\bf #1}, #2 (#3)}
\def\npbfsold#1#2#3#4{Nucl.\ Phys.\ {\bf #1} [FS #2], #3, (#4)}
\def\npbfs#1#2#3{Nucl.\ Phys.\ {\bf #1} [FS], #2, (#3)}
\def\plb#1#2#3{Phys.\ Lett.\ B {\bf #1}, #2 (#3)}
\def\physrep#1#2#3{Phys.\ Rep.\ {\bf #1}, #2 (#3)}
\def\advphys#1#2#3{Adv.\ in Phys.\ {\bf #1}, #2 (#3)}
\def\mpla#1#2#3{Mod.\ Phys.\ Lett.\ A {\bf #1}, #2 (#3)}
\def\mplb#1#2#3{Mod.\ Phys.\ Lett.\ B {\bf #1}, #2 (#3)}
\def\ijmpa#1#2#3{Int.\ J.\ Mod.\ Phys.\ A {\bf #1}, #2 (#3)}
\def\ijmpb#1#2#3{Int.\ J.\ Mod.\ Phys.\ B {\bf #1}, #2 (#3)}
\def\rmp#1#2#3{Rev.\ Mod.\ Phys.\ {\bf #1}, #2 (#3)}
\def\jpc#1#2#3{J.\ Phys.\ C {\bf #1}, #2 (#3)}
\def\jpa#1#2#3{J.\ Phys.\ A {\bf #1}, #2 (#3)}
\def\physicac#1#2#3{Physica C {\bf #1}, #2 (#3)}
\def\physicaa#1#2#3{Physica A {\bf #1}, #2 (#3)}
\def\physicab#1#2#3{Physica B {\bf #1}, #2 (#3)}
\def\physicae#1#2#3{Physica E {\bf #1}, #2 (#3)}
\def\nature#1#2#3{Nature {\bf #1}, #2 (#3)}
\def\science#1#2#3{Science {\bf #1}, #2 (#3)}
\def\bams#1#2#3{Bull. Am. Math. Soc. {\bf #1}, #2 (#3)}
\def\baps#1#2#3{Bull. Am. Phys. Soc. {\bf #1}, #2 (#3)}
\def\cmp#1#2#3{Comm.\ Math.\ Phys.\ {\bf #1}, #2 (#3)}
\def\jmp#1#2#3{J. Math. Phys. {\bf #1}, #2 (#3)}
\def\jhep#1#2#3{J. High Ener. Phys. {\bf #1}, #2 (#3)}
\def\jstatphys#1#2#3{J. Stat. Phys. {\bf #1}, #2 (#3)}
\def\annphys#1#2#3{Annals Phys.\ {\bf #1}, #2 (#3)}

\end{document}